\documentclass[11pt,a4paper]{article}
\usepackage{cite}
\usepackage{url}
\usepackage{amsfonts}
\usepackage{amsmath}
\usepackage{amssymb}
\usepackage{graphicx}
\usepackage{epsfig}
\usepackage{lineno}
\usepackage[usenames]{color}
\usepackage{hyperref}
\usepackage{subfigure}
\usepackage[normalem]{ulem} 
\usepackage{color}
\usepackage{units}
\usepackage{authblk}
\usepackage[titletoc,page]{appendix}

\newcommand{\com}[1]       {{\relax}}

\graphicspath{{./img/}}

\begin{document}
\title{Correction methods for finite-acceptance effects \\ in two-particle correlation analyses}
\author{Saehanseul Oh}
\author{Tim Schuster}
\affil{\small Yale University, New Haven, CT, 06511, USA}
\author{Andreas Morsch}
\affil{\small CERN, 1211 Geneva 23, Switzerland}
\author{Constantin Loizides}
\affil{\small LBNL, Berkeley, CA, 94720, USA}
\date{}               

\maketitle
\vspace{-1.cm}
\begin{abstract}
Two-particle angular correlations have been widely used as a tool to explore particle production mechanisms in heavy-ion collisions. The mixed-event technique is generally used as a standard method to correct for finite-acceptance effects. We demonstrate that event mixing only provides an approximate acceptance correction, and propose new methods for finite-acceptance corrections. Starting from discussions about 2-dimensional correction procedures, new methods are derived for specific assumptions on the properties of the signal, such as uniform signal distribution or $\delta$-function-like trigger particle distribution, and suitable for two-particle correlation analyses from particles at mid-rapidity and jet-hadron or high $p_{\text{T}}$-triggered hadron-hadron correlations. Per-trigger associated particle yields from the mixed-event method and the new methods are compared through Monte Carlo simulations containing well-defined correlation signals. Significant differences are observed at large pseudorapidity differences in general and especially for asymmetric particle distribution like that produced in proton--nucleus collisions. The applicability and validity of the new methods are discussed in detail.
\end{abstract}

\enlargethispage{2cm}
\makeatletter
\renewcommand\tableofcontents{%
    \@starttoc{toc}%
}
\makeatother
\vspace{0.25cm}
{
\noindent {\bf Contents}

\footnotesize 
\tableofcontents
}

\thispagestyle{empty} 
\clearpage

\section{Introduction}
Two-particle angular correlations have been widely used in the field of relativistic heavy-ion physics to provide information on particle production mechanisms in various collisional systems. 
Away-side jet suppression and collective flow, both distinctive features of nuclear collisions, have been observed through two-particle correlations at RHIC~\cite{STAR:2PC,PHENIX:Flow}. 
At the LHC, the observation of a near-side ridge structure in pp~\cite{Khachatryan:2010gv} and the discovery of a double-ridge structure in p--Pb collisions~\cite{ALICE:2PC:pPb,Aad:2012gla,CMS:2012qk} have opened a new debate on the origin of these structures.
Throughout, event-mixing procedures have been considered as a standard technique for the pair acceptance correction in two-particle correlation analysis. 
However, a few recent papers have pointed out shortcomings of the conventional correction with event-mixing, and proposed new methods using single-particle efficiency$\times$acceptance functions~\cite{Method:Fuqiang} and multi-dimensional weights~\cite{Method:Pruneau:2}.

In this article, finite-acceptance effects ---as distinguished from detector efficiency effects--- in two-particle angular correlation analyses are discussed in detail, and alternative correction methods for finite-acceptance effects are derived and tested. 
The derivations are obtained by comparison to the ideal case without finite-acceptance effects, which ensures mathematical completeness, assuming translational invariance of the signal. 

Angular correlation studies involve measuring the distributions of the relative azimuthal angle $\Delta \varphi$ or relative pseudo-rapidity $\Delta \eta$ between particle pairs consisting of a trigger particle in a certain transverse momentum, $p_{\rm T, trig}$, interval and an associated particle in a $p_{\rm T, assoc}$ interval. 
For $x$ being the coordinate with finite acceptance, a general correction method in $(x_{\text{t}},\,  \Delta x)$ space is first discussed, where $x_{\text{t}}$ and $\Delta x$ correspond to the trigger particle $x$ and difference between trigger particle and associated particle $x$, respectively. 
New methods in $\Delta x$ space are obtained by making assumptions on the properties of the signal:
uniform signal distribution in $x$ and, for jet-like correlations, a $\delta$-function-like distribution of the trigger particle with respect to the jet axis. 
The validity of these methods depends on the similarity of the signal characteristics in data to the assumed conditions. 
The method assuming a uniform signal distribution is suitable for the correlation analyses using particles measured at midrapidity in symmetric nucleus-nucleus collisions. The method assuming a $\delta$-function-like trigger particle distribution is suitable for the study of near-side jet--hadron or high $p_{\text{T}}$-triggered hadron--hadron correlations. 
While the correction with the event-mixing technique is equivalent to producing a normalized ratio function of correlated and uncorrelated particle production, the new methods are independent of the uncorrelated background.

We apply the new methods to Monte Carlo simulations to test their validity, and point out problems of the conventional mixed-event technique.
The Monte Carlo simulations contain well-defined correlation signals: fragmentation of a dijets generated with PYTHIA event generator or a global correlation of all partices with a common symmetry plane in a toy model.

The article is organized as follows: Section \ref{Sec:Def} defines the mathematical notation of correlation function and per-trigger normalized associated particle yield, and describes finite-acceptance effects in two-particle correlation analysis. The new methods are introduced in Section \ref{Sec:AM}, and tested with Monte Carlo simulation in Section \ref{Sec:MC}. In Section \ref{Sec:Sum}, we summarize our results. Appendix \ref{Sec:App} provides details on the derivation of the new methods.

\section{Definitions}
\label{Sec:Def}
\subsection{Correlation function and per-trigger yield}
Two-particle correlation studies are based on the simultaneous measurement of pairs of particles in each event. The results might be affected by various particle production and transport processes, such as radial flow, elliptic flow, resonance decays, jets and others~\cite{Method:Pruneau:1}. Mathematically, trigger and associated single-particle densities are denoted as functions of azimuthal angle and pseudorapidity, $\rho_{\text{t}}(\varphi_{\text{t}}, \eta_{\text{t}})$ and $\rho_{\text{a}}(\varphi_{\text{a}}, \eta_{\text{a}})$, where the subscript ``t''~(``a'') stands for trigger~(associated) particles. Following the same notation, the two-particle density of trigger and associated particle pairs is denoted by $\rho_{\text{a,t}}(\varphi_{\text{a}}, \eta_{\text{a}};\varphi_{\text{t}},\eta_{\text{t}})$~\cite{Method:Pruneau:2,Method:Vechernin}. Trigger and associated particles are most commonly selected by either transverse momentum ($p_{\text{T}}$) or particle species. The density functions are defined as
\begin{eqnarray}
\rho_{\text{t}}(\varphi_{\text{t}}, \eta_{\text{t}})=\frac{\text{d}^2N_{\text{t}}}{\text{d}\varphi_{\text{t}}\,\text{d}\eta_{\text{t}}} \; \text{,} \qquad \rho_{\text{a}}(\varphi_{\text{a}}, \eta_{\text{a}})=\frac{\text{d}^2N_{\text{a}}}{\text{d}\varphi_{\text{a}}\,\text{d}\eta_{\text{a}}}\; \text{,} \nonumber \\
\rho_{\text{a,t}}(\varphi_{\text{a}}, \eta_{\text{a}};\varphi_{\text{t}}, \eta_{\text{t}})=\frac{\text{d}^4N_{\text{a,t}}}{\text{d}\varphi_{\text{a}}\,\text{d}\eta_{\text{a}}\;\text{d}\varphi_{\text{t}}\,\text{d}\eta_{\text{t}}} \; \text{.}
\end{eqnarray}
Typically, the definition of correlation function, $C_{2,\text{R}}(\eta_{\text{a}}\,\eta_{\text{t}};\varphi_{\text{a}}\,\varphi_{\text{t}})$, is
\begin{eqnarray}
C_{2,\text{R}}(\varphi_{\text{a}},\varphi_{\text{t}};\eta_{\text{a}},\eta_{\text{t}}) &=& \frac{\rho_{\text{a,t}}(\varphi_{\text{a}}, \eta_{\text{a}};\varphi_{\text{t}}, \eta_{\text{t}})-\rho_{\text{a}}(\varphi_{\text{a}}, \eta_{\text{a}})\,\rho_{\text{t}}(\varphi_{\text{t}}, \eta_{\text{t}})}{\rho_{\text{a}}(\varphi_{\text{a}}, \eta_{\text{a}})\,\rho_{\text{t}}(\varphi_{\text{t}}, \eta_{\text{t}})}\nonumber\\
&=&\frac{\rho_{\text{a,t}}(\varphi_{\text{a}}, \eta_{\text{a}};\varphi_{\text{t}}, \eta_{\text{t}})}{\rho_{\text{a}}(\varphi_{\text{a}}, \eta_{\text{a}})\,\rho_{\text{t}}(\varphi_{\text{t}}, \eta_{\text{t}})}-1\; \text{,}
\end{eqnarray}
where ``R'' stands for ratio.
Assuming rotational invariance in azimuth, one can write
\begin{eqnarray}
\rho_{\text{t}}(\varphi_{\text{t}}, \eta_{\text{t}})=\frac{\rho_{\text{t}}(\eta_{\text{t}})}{2\pi} \; \text{,} \qquad \rho_{\text{a}}(\varphi_{\text{a}}, \eta_{\text{a}})=\frac{\rho_{\text{a}}(\eta_{\text{a}})}{2\pi} \; \text{,} \nonumber\\
\rho_{\text{a,t}}(\varphi_{\text{a}}, \eta_{\text{a}};\varphi_{\text{t}}, \eta_{\text{t}}) = \frac{\rho_{\text{a,t}}(\varphi_{\text{t}}-\varphi_{\text{a}}; \eta_{\text{a}}, \eta_{\text{t}})}{(2\pi)^{2}} \; \text{,}
\end{eqnarray}
and
\begin{eqnarray}
C_{2,\text{R}}(\varphi_{\text{a}},\varphi_{\text{t}};\eta_{\text{a}},\eta_{\text{t}}) &=& C_{2,\text{R}}(\varphi_{\text{t}}-\varphi_{\text{a}}; \eta_{\text{a}},\eta_{\text{t}}) \nonumber \\
&=& \frac{\rho_{\text{a,t}}(\varphi_{\text{t}}-\varphi_{\text{a}}; \eta_{\text{a}}, \eta_{\text{t}})}{\rho_{\text{a}}(\eta_{\text{a}})\;\rho_{\text{t}}(\eta_{\text{t}})} -1 \; \text{.}
\end{eqnarray}
Experimentally, a simpler correlation function is often used. The correlation function $C_{\text{R}}(\Delta\varphi, \Delta\eta)$ with $\Delta\varphi = \varphi_{\text{t}} - \varphi_{\text{a}}$ and $\Delta\eta = \eta_{\text{t}} - \eta_{\text{a}}$ is defined by
\begin{equation}
C_{\text{R}}(\Delta\varphi, \Delta\eta)= \frac{S(\Delta\varphi, \Delta\eta)}{B(\Delta\varphi, \Delta\eta)}-1\;\text{,}
\label{CrEq}
\end{equation}
where
\begin{equation}
S(\Delta\varphi, \Delta\eta)=\frac{\text{d}^2N_{\text{pair}}}{\text{d}\Delta\varphi\,\text{d}\Delta\eta}
\end{equation}
is the two-particle distribution in the same events, while $B(\Delta\varphi, \Delta\eta)$ is constructed from two particles from different events and corresponds to uncorrelated particle production. 
The definition of $C_{\text{R}}(\Delta\varphi, \Delta\eta)$ can be interpreted to inherently assume the translational invariance of the correlated signal in $\eta$ in addition to $\varphi$, or as an average of $\Delta\eta$ structure within the considered $\eta$ range. 
In general, this involves loss of information such as the $\eta_{\text{t}}$ dependence, but $C_{\text{R}}$ provides a simpler representation of the correlation. 
Using $(\eta_{\text{a}},\eta_{\text{t}})$ instead of $\Delta\eta$ is suggested in \cite{Method:Pruneau:2}, and used \textit{e.g.}\ in \cite{Alver:2010rt}.
Instead, the present article focuses on the use of $\Delta\eta$, which is technically simpler and has a statistical benefit compared to the one based on $(\eta_{\text{a}}, \eta_{\text{t}})$.

Another correlation observable~\cite{STAR:2005ph,PHENIX:2005ee,PHENIX:2008ae,CMS:2PC} used instead of the above is the per-trigger normalized associated particle yield (per-trigger yield),
\begin{equation}
C_{\text{yield}}(\Delta\varphi, \Delta\eta)=\frac{1}{N_{\text{trig}}}\frac{\text{d}^2N_{\text{pair}}}{\text{d}\Delta\phi\,\text{d}\Delta\eta}\;\text{,}
\label{CptrigY}
\end{equation}
and it is generally approximated by
\begin{equation}
\frac{1}{N_{\text{trig}}}\frac{\text{d}^2N_{\text{pair}}}{\text{d}\Delta\phi\,\text{d}\Delta\eta} \simeq C_{\text{trig,R}}=B(0,0)\,\frac{S(\Delta\varphi,\,\Delta\eta)}{B(\Delta\varphi,\,\Delta\eta)}
\label{CtrigR}
\end{equation}
where
\begin{equation}
S(\Delta\varphi,\,\Delta\eta)=\frac{1}{N_{\text{trig}}^{\text{same}}}\frac{\text{d}^2N^{\text{\text{same}}}}{\text{d}\Delta\varphi\,\text{d}\Delta\eta}\;\text{,}\qquad B(\Delta\varphi,\,\Delta\eta)=\frac{1}{N_{\text{trig}}^{\text{mixed}}}\frac{\text{d}^2 N^{\text{\text{mixed}}}}{\text{d}\Delta\varphi\,\text{d}\Delta\eta}\;\text{,}
\end{equation}
within specific $\eta$-acceptance ranges. 

Essentially, $C_{2,\text{R}}$, $C_{\text{R}}$, and $C_{\text{trig,R}}$ contain the same information, namely how much the correlated production differs from the uncorrelated production, and they depend on a ratio between these two. 
This is indicated by the common index ``R''. 
In $C_{\text{trig,R}}$, it is often assumed that the division by $B(\Delta\varphi, \Delta\eta)$ represents a correction for pair-acceptance effects~\cite{STAR:2005ph,PHENIX:2005ee,PHENIX:2008ae,ALICE:2PC:pPb,CMS:2012qk}. 
However, we claim that $C_{\text{trig,R}}$ is not corrected properly for the finite-acceptance effects, and the normalized ratio function is only an approximation of the intended per-trigger yield. 
As will be discussed in more detail throughout the article, the extraction of yields or $\Delta\varphi$-projections are affected by distortions inherent to $C_{\text{trig,R}}$. 
In the following, $C$ represents the per-trigger yield, not defined by a ratio, and exact correction methods for finite-acceptance effects will be discussed.

\subsection{Finite-acceptance effects}
\label{sec:FAeffects}
Finite-acceptance effects in a two-particle correlation analysis should be distinguished from detector efficiency effects. 
In recent correlation analyses, single-particle efficiency effects are corrected by the corresponding efficiency obtained with event generators followed by  detector simulations at the moment when $(\Delta\varphi, \Delta\eta)$ bins of per-trigger yields are filled~\cite{ALICE:2PC:pPb,Aad:2012gla,CMS:2012qk}. 
What is done to correct for finite-acceptance effects is to divide the same-event function by the normalized mixed-event function, producing a ratio function as described in the previous section. 
It is easier to demonstrate the complication of finite-acceptance effects in two-particle correlations with a 1-dimensional example. 
Considering a case where the detector acceptance range is $[a_{1}, a_{2}]$ in $x$, we assume translational invariance of the correlated signal in $x$, which makes it possible to use $\Delta x=x_{\text{t}}-x_{\text{a}}$, instead of $(x_{\text{a}}, x_{\text{t}})$. 
Then $X$, $f_{\text{t}}(x-X)$, $f_{\text{a}}(x-X)$, and $g(X)$ are defined as
\begin{eqnarray}
X &\equiv& \text{Common reference point of the trigger and associated particle distributions}  \nonumber \\
&&\text{for each correlated signal, where $X$ denotes the center of trigger and associated} \nonumber \\
&&\text{particle distributions for convenience;} \nonumber \\
f_{\text{t}}(x-X) &\equiv& \text{Trigger particle distribution in a correlated signal with respect to } x=X \text{ in}\nonumber \\
&&\text{the range } -b < x-X < b\text{, where } 2b \text{ corresponds to the size of a correlated}\nonumber \\
&&\text{signal in } x \text{;}\nonumber \\
f_{\text{a}}(x-X) &\equiv& \text{Associated particle distribution in a correlated signal with respect to } x=X \nonumber \\
&&\text{ in the range } -b < x-X < b\text{;} \nonumber \\
g(X) &\equiv& \text{Distribution of }X\text{ over all events.}\nonumber 
\end{eqnarray}
The definitions assume that the trigger and associated particle distributions have a common reference point, $X$, in $x$ for each correlated signal in addition to the translational invariance. For example, this is most relevant to particle correlations in a jet and the reference point corresponds to the jet axis. 
In the definitions, $b$ can be an arbitrarily large value as far as it includes the correlated signal, as dependence on $b$ will be eliminated in the new methods.
For infinite acceptance in $x$, the per-trigger yield from a single correlated signal for a given $X$, $C_{\text{inf,single}}$, with $f_{\text{a}}$ and $f_{\text{t}}$ is mathematically defined by the cross-correlation\footnote{Cross-correlation: $(f\star g)(\Delta x) = \int f(x - \Delta x)\,g(x)\,\text{d}x$},
\begin{eqnarray}
C_{\text{inf,single}}(\Delta x) &=& \frac{1}{(N_{\text{trig}})_{\text{inf,single}}}(f_{\text{a}}\star f_{\text{t}})_{\text{inf,single}} \nonumber \\
&=& \frac{1}{(N_{\text{trig}})_{\text{inf,single}}}\int_{-\infty}^\infty f_{\text{a}}(x-X-\Delta x)\,f_{\text{t}}(x-X)\,\text{d}x \nonumber\\
&=& \frac{1}{(N_{\text{trig}})_{\text{inf,single}}}\int_{\max(X-b, X-b+\Delta x)}^{\min(X+b, X+b+\Delta x)}f_{\text{a}}(x-X-\Delta x)\,f_{\text{t}}(x-X)\,\text{d}x \nonumber \\
&=& \frac{1}{(N_{\text{trig}})_{\text{inf,single}}}\int_{\max(-b, -b+\Delta x)}^{\min(b, b+\Delta x)}f_{\text{a}}(x'-\Delta x)\,f_{\text{t}}(x')\,\text{d}x' \;\text{,} 
\label{CidealSingle}
\end{eqnarray}
where
\begin{eqnarray}
(N_{\text{trig}})_{\text{inf,single}} &=& \int_{-\infty}^{\infty} f_{\text{t}}(x-X)\,\text{d}x = \int_{X-b}^{X+b} f_{\text{t}}(x-X)\,\text{d}x \nonumber \\
&=& \int_{-b}^{b} f_{\text{t}}(x')\,\text{d}x'\;\text{.}
\end{eqnarray}
We note that $(f_{\text{a}}\star f_{\text{t}})_{\text{inf,single}}$ and $(N_{\text{trig}})_{\text{inf,single}}$ do not depend on $X$, as expected from the assumption of the translational invariance. With infinite acceptance, the per-trigger yield over all events with all signals should be the same as the above $C_{\text{inf,single}}(\Delta x)$ as
\begin{eqnarray}
(N_{\text{trig}})_{\text{inf}} &=& \int_{-\infty}^{\infty}g(X)\,\left(\int_{-\infty}^{\infty} f_{\text{t}}(x-X)\,\text{d}x\right)\,\text{d}X \nonumber \\
& = & (N_{\text{trig}})_{\text{inf,single}}\int_{-\infty}^{\infty}g(X)\,\text{d}X  \;\text{,}
\label{Nideal}
\end{eqnarray}
\begin{eqnarray}
C_{\text{inf}}(\Delta x) &=& \frac{1}{(N_{\text{trig}})_{\text{inf}}}\int_{-\infty}^{\infty}g(X)\,\left(\int_{-\infty}^\infty f_{\text{a}}(x-X-\Delta x)\,f_{\text{t}}(x-X)\,\text{d}x\right)\,\text{d}X \nonumber \\
&=&  \frac{1}{(N_{\text{trig}})_{\text{inf,single}}\int_{-\infty}^{\infty}g(X)\,\text{d}X}\int_{-\infty}^{\infty}g(X)\,\text{d}X\left(\int_{-\infty}^\infty f_{\text{a}}(x-X-\Delta x)\,f_{\text{t}}(x-X)\,\text{d}x\right)\nonumber \\
&=& \frac{1}{(N_{\text{trig}})_{\text{inf,single}}}\int_{-\infty}^\infty f_{\text{a}}(x-X-\Delta x)\,f_{\text{t}}(x-X)\,\text{d}x \nonumber\\
&=& C_{\text{inf,single}}(\Delta x)\; \text{.}
\label{Cideal}
\end{eqnarray}

If we consider the case for finite acceptance, $[a_{1}, a_{2}]$ in $x$, parts of $f_{\text{t}}$ and $f_{\text{a}}$ cannot be detected depending on $X$ of the signal. For instance, for a signal with $X=a_{1}$, only the positive part of $f_{\text{t}}$ and $f_{\text{a}}$ (where $x-X>0$) can contribute to the per-trigger yield. If the whole range of the correlated signal is within $[a_{1}, a_{2}]$, no finite-acceptance effects are involved. Finite-acceptance effects mean that pairs are not counted depending on the correlated particle positions, and must be corrected for. The per-trigger yield in a single event with given $X$ and finite acceptance can be written as
\begin{equation}
C_{\text{single}}(\Delta x)= \frac{1}{(N_{\text{trig}})_{\text{single}}}\int_{-\infty}^\infty f_{\text{a}}(x-X-\Delta x)\,A_{\text{a}}(x-\Delta x)\,f_{\text{t}}(x-X)\,A_{\text{t}}(x)\,\text{d}x \; \text{,}
\label{1DCrEq_single}
\end{equation}
where we have introduced an acceptance operator
\begin{eqnarray}
  A_{\text{t}}(x) =
  \begin{cases}
   1 & \text{if } a_{1\text{,t}} < x < a_{2\text{,t}} \\
   0 & \text{otherwise }
  \end{cases}\;\text{,}\qquad
  A_{\text{a}}(x) =
  \begin{cases}
   1 & \text{if } a_{1\text{,a}} < x < a_{2\text{,a}} \\
   0 & \text{otherwise }
  \end{cases}\;\text{,}
  \label{AccEq}
\end{eqnarray}
and
\begin{equation}
(N_{\text{trig}})_{\text{single}}=\int_{-\infty}^{\infty} f_{\text{t}}(x-X)\,A_{\text{t}}(x)\,\text{d}x \;\text{.}
\end{equation}
By treating separately $A_{\text{a}}$ and $A_{\text{t}}$ as Eq.~(\ref{AccEq}), following formalism can be applied to cases where the correlation between different $x$ ranges of trigger and associated particles is studied~\cite{Adamczyk:2015qy,PHENIX:2006hi}. 
Then considering all events,
\begin{equation}
C(\Delta x) = \frac{1}{N_\text{trig}}\int_{-\infty}^{\infty}\int_{-\infty}^\infty g(X)\,f_{\text{a}}(x-X-\Delta x)\,A_{\text{a}}(x-\Delta x)\,f_{\text{t}}(x-X)\,A_{\text{t}}(x)\,\text{d}x\,\text{d}X\;\text{,}
\label{1DCrEq}
\end{equation}
\begin{equation}
N_{\text{trig}}=\int_{-\infty}^{\infty}\int_{-\infty}^{\infty} g(X)\,f_{\text{t}}(x-X)\,A_{\text{t}}(x)\,\text{d}x\,\text{d}X\;\text{.}
\label{1DNtrig}
\end{equation}

There is no general formula to relate $C_{\text{inf}}(\Delta x)$ and $C(\Delta x)$, as the integrand of Eq.~(\ref{Cideal}) is multiplied by acceptance operators in Eq.~(\ref{1DCrEq}). 
In the following section, we will discuss the correction of finite-acceptance effects at $(x, \Delta x)$ space before the integration in Eq.~(\ref{1DCrEq}) is performed. 
Also, exact formulas which connect $C_{\text{inf}}(\Delta x)$ and $C(\Delta x)$ will be derived under specific assumptions of the correlated signal. 
These formulas might be used as approximations in realistic cases to correct for finite-acceptance effects.

\section{Alternative methods}
\label{Sec:AM}
Without loss of generality we restrict the following discussion of the correction for finite-acceptance effects to measurements as a function of one variable, $x$. 
Actual measurements are performed typically as a function of $\Delta\varphi$ and $\Delta\eta$. 
Note that because of azimuthal symmetry, all assumptions made in the following about the dependence of the signal on $x$ and $\Delta x$ are realized for $x=\varphi$ and the corrections are exact in case of full azimuthal acceptance, whereas for $x=\eta$, they are approximate. 
Since these correction methods also preserve the additivity of signal and uncorrelated background, it is irrelevant whether one performs the background subtraction~\cite{STAR:2005ph,PHENIX:2005ee,PHENIX:2008ae,Aggarwal:2010rf,ALICE:2PC:pPb,CMS:2PC,Ajitanand:2005jj,Trainor:2009gj} before or after the acceptance correction. 
Hence, we can also restrict our discussion to the correction of the signal only.

\subsection{Correction methods and scope of applicability\label{ch3-1}}
$C_{\text{single}}(\Delta x)$ in Eq.~(\ref{1DCrEq_single}) can be considered as a weighted average of $C_{\text{single}}(x_{\text{t}}, \Delta x)$, per-trigger yield in $(x_{\text{t}}, \Delta x)$ space, over $x_{\text{t}}$ as $x$ in the integral corresponds to the $x$ of trigger particles. 
The integrand is equivalent to $C_{\text{single}}(x_{\text{t}}, \Delta x)\,N_{\text{trig,single}}(x_{\text{t}})$, where $N_{\text{trig,single}}(x_{\text{t}}) = f_{\text{t}}(x_{\text{t}}-X)\,A_{\text{t}}(x_{\text{t}})$ corresponds to the weight of the average. 
$C_{\text{single}}(x_{\text{t}}, \Delta x)$ and its weighted average over $x_{\text{t}}$ for given $X$ can be free of finite-acceptance effects under certain conditions, such as trigger (associated) particle distribution being within the trigger (associated) particle acceptance, larger associated particle acceptance range than trigger particle acceptance range, and more generally when
\begin{equation}
a_{2\text{,t}} - a_{2\text{,a}}< \Delta x < a_{1\text{,t}} - a_{1\text{,a}}\;\text{.}
\end{equation}
One realistic example is near-side per-trigger yield from jet-hadron correlation, with trigger particle acceptance $[-\frac{a}{2}, \frac{a}{2}]$ and associated particle acceptance $[-a, a]$. 
In this case, both $C_{\text{single}}(x_{\text{t}}, \Delta x)$ and $C_{\text{single}}(\Delta x)$ do not experience finite-acceptance effects for $\Delta x$ within $[-\frac{a}{2}, \frac{a}{2}]$.

Eq.~(\ref{1DCrEq}) is the weighted average of $C_{\text{single}}(x_{\text{t}}, \Delta x)$ over $x_{\text{t}}$ and $X$, and equivalent to the weighted average of single per-trigger yields, $C_{\text{single}}(\Delta x)$, over $X$ where the yield is weighted with $g(X)$. 
As $C_{\text{single}}(x_{\text{t}}, \Delta x)$ is free of finite-acceptance effects with certain conditions, $C(\Delta x)$ can also be free of them when the average over $x_{\text{t}}$ and $X$ is selectively performed with only $C_{\text{single}}(x_{\text{t}}, \Delta x)$ with no finite-acceptance effects. 
Although restricting trigger-particle acceptance range may cause lower statistics in real analysis, this certainly is one way to avoid finite-acceptance effects in two-particle correlation analysis. 
An additional benefit of considering per-trigger yield in $(x_{\text{t}}, \Delta x)$-space is the possibility that the assumption of translational invariance of the signal can be explicitly checked. 

If one can find a form that can relate $C_{\text{inf}}(\Delta x)$ and $C(\Delta x)$, this form is simply interpreted as a correction form of finite-acceptance effects. 
But more generally, the correction procedure is interpreted as restoring the shape of two-particle correlation signal with finite-acceptance effects at $(x_{\text{t}}, \Delta x)$ space and averaging the corrected $C(x_{\text{t}}, \Delta x)$ over $x_{\text{t}}$. 
In the formulation of the correction form of finite-acceptance effects, comparing $C(\Delta x)$ with $C_{\text{inf}}(\Delta x)$ automatically assumes the restoration of distorted correlation signal shapes from finite-acceptance effects. 
Meanwhile, directly applying the second interpretation, one can consider the correction in $(x_{\text{t}}, \Delta x)$ space by copying the shape of correlated signal without finite-acceptance effects within certain $x_{\text{t}}$ and ${\Delta x}$ ranges into the distorted signals with finite-acceptance effects in other $x_{\text{t}}$, assuming translational invariance of the signal.
After the correction in $(x_{\text{t}}, \Delta x)$ space, corrected $C(\Delta x)$ can be estimated by averaging $C(x_{\text{t}}, \Delta x)$ over $x_{\text{t}}$ with weight, $N_{\text{trig}}(x_{\text{t}})$.   

In addition to the general correction procedures in $(x_{\text{t}}, \Delta x)$ space, the correction factor can be factorized out under certain conditions and possibly performed after averaging $C(x_{\text{t}}, \Delta x)$ over $x_{\text{t}}$ without correction for the finite-acceptance effects. 
In this case, one does not have to generate $C(x_{\text{t}}, \Delta x)$ and consider $x_{\text{t}}$ and $\Delta x$ ranges, which should be used as the bases of the restoration of the distorted signals, but only needs to generate uncorrected $C(\Delta x)$. 
Continuing from Eq.~(\ref{1DCrEq}), where there is a single common reference point between trigger and associated particle distributions such as near-side ($\Delta\varphi\approx0$) jet-like correlations, we can find two simple relations between $C_{\text{inf}}$ and $C$ under specific assumptions. Detailed derivations can be found in Appendix \ref{Sec:App}, and only the final formulas and their applicability will be given and discussed in the current section. If the distribution of the signal, $g(X)$, is constant, then
\begin{eqnarray}
C(\Delta x) =  \frac{(A_{\text{a}}\star A_{\text{t}})(\Delta x)}{\Delta_{\text{t}}}\,C_{\text{inf}}(\Delta x)\; \text{,}
\label{Method2}
\end{eqnarray}
where $\Delta_{\text{t}}=a_{2\text{,t}}-a_{1\text{,t}}$ is the size of trigger particle acceptance. In other words, the measured per-trigger yield in case of constant signal distribution can be corrected back to the case without finite-acceptance effects by dividing by $\frac{(A_{\text{a}}\star A_{\text{t}})(\Delta x)}{\Delta_{\text{t}}}$. 

Another condition when the exact formula can be derived is when $f_{\text{t}}$ is a $\delta$-function. This is similar to the near-side jet-hadron correlations in a jet or high $p_{\text{T}}$-triggered hadron-hadron correlations. In this case, 
\begin{eqnarray}
C(\Delta x) = \frac{(A_{\text{a}}\star n_{\text{trig}}A_{\text{t}})(\Delta x)}{N_{\text{trig}}}\,C_{\text{inf}}(\Delta x)\; \text{,}
\label{Method3}
\end{eqnarray}
where $n_{\text{trig}}(x)$ is a measured trigger particle distribution in $x$ within the acceptance. As a result, we can correct the measured per-trigger yield back to the per-trigger yield with infinite acceptance by dividing by $\frac{(A_{\text{a}}\star n_{\text{trig}}A_{\text{t}})(\Delta x)}{N_{\text{trig}}}$.

Besides a single common reference point case as discussed up to now, we can also consider a different type of correlation signal, which has two reference points between trigger and associated particle distributions in each event. This case is most relevant to the particle correlations in back-to-back di-jet events where each jet represents a reference point. If two reference points are $X$ and $Y$, the correlation formula between trigger and associated particle with respect to the same reference point $X$($Y$) has already been derived. However, the correlation between the trigger particle distribution with respect to the reference point $X$($Y$) and the associated particle distribution with respect to the reference point $Y$($X$) should be dealt with a different way. In di-jet events, these cross terms correspond to the away-side structure near $\Delta\varphi = \pi$, and are distinguished from correlations in a single jet. If we assume a case with infinite acceptance, 
\begin{eqnarray}
C_{\text{inf}}(\Delta x) &=& \frac{1}{N_\text{trig,inf}}\int_{-\infty}^{\infty}\text{d}x\int_{-\infty}^{\infty}\text{d}Y\int_{-\infty}^{\infty}\text{d}X\,g(X,Y)\Big( f_{\text{1,a}}(x-Y-\Delta x)\,f_{\text{2,t}}(x-X)\nonumber \\
&&+f_{\text{1,t}}(x-Y)\,f_{\text{2,a}}(x-X-\Delta x)\Big) \text{,}
\end{eqnarray}
\begin{eqnarray}
N_\text{trig,inf} = \int_{-\infty}^{\infty}\text{d}x\int_{-\infty}^{\infty}\text{d}X\int_{-\infty}^{\infty}\text{d}Y\,g(X,Y)\,\Big(f_{\text{1,t}}(x-Y) + f_{\text{2,t}}(x-X)\Big)\text{,}
\end{eqnarray}
where $g(X,Y)$ represents the distribution of two reference points over all events. With finite acceptance as defined in Eq.~(\ref{AccEq}),
\begin{eqnarray}
C(\Delta x) &=& \frac{1}{N_\text{trig}}\left(\int_{-\infty}^{\infty}\text{d}x\int_{-\infty}^{\infty}\text{d}Y\int_{-\infty}^{\infty}\text{d}X\,g(X,Y)\,A_{\text{a}}(x-\Delta x)\,A_{\text{t}}(x) \Big( \right.\nonumber \\
&&\left.f_{\text{1,a}}(x-Y-\Delta x)\,f_{\text{2,t}}(x-X)\,+ f_{\text{1,t}}(x-Y)\,f_{\text{2,a}}(x-X-\Delta x)\,\Big)\right) \text{,}
\label{2RefCtrig}
\end{eqnarray}
\begin{eqnarray}
N_\text{trig} = \int_{-\infty}^{\infty}\text{d}x\int_{-\infty}^{\infty}\text{d}X\int_{-\infty}^{\infty}dY\,g(X,Y)\,A_{\text{t}}(x)\Big(f_{\text{1,t}}(x-Y) + f_{\text{2,t}}(x-X)\Big)\text{.}
\label{2RefNtrig}
\end{eqnarray}
Like the derivation for a single common reference correlation, there is no general formula by which we can relate the measured per-trigger yield to the per-trigger yield without finite-acceptance effects. However, an approximate relation can be found in simple cases. First of all, it is reasonable to assume that $f_{\text{1,t}} = f_{\text{2,t}}=f_{\text{t}}$ and $f_{\text{1,a}} = f_{\text{2,a}}=f_{\text{a}}$. Since two reference points are indistinguishable, we know that $g(X,Y)=g(Y,X)$. Then $g(X,Y)$ can be rewritten as $g(X-Y, X+Y)$, and the simplest assumption for $g(X-Y,X+Y)$ is that this function only depends on the distance between $X$ and $Y$, which is $|X-Y|$. This formula includes constant $g(X,Y)$ case. Then we can write,
\begin{equation}
g(X-Y,X+Y) = G(|X-Y|) = G(X-Y)\;\text{,}
\label{2RefCond}
\end{equation}
as $G(X-Y) = G(Y-X)$, and assume without loss of generality that $G(X-Y)$ has only values if $-c<X-Y<c$ with sufficiently large $c$ compared to $2b$, the range of $f_{\text{t}}$ and $f_{\text{a}}$. Then
\begin{eqnarray}
C(\Delta x) =  \frac{(A_{\text{a}}\star A_{\text{t}})(\Delta x)}{\Delta_{\text{t}}}\,C_{\text{inf}}(\Delta x)\; \text{.}
\label{Method2A}
\end{eqnarray}
This is the same result as for the constant $g(X)$ case in the single common reference correlation, Eq.~(\ref{Method2}). Also, if 
\begin{equation}
g(X,Y) = h(X)F(X-Y)\;\text{,}
\label{2RefBias}
\end{equation}
with $g(X,Y) = g(Y,X)$ and $f_{\text{t}}$ is a $\delta$-function as in jet-hadron correlations, 
\begin{eqnarray}
C(\Delta x) = \frac{(A_{\text{a}}\star n_{\text{trig}}A_{\text{t}})(\Delta x)}{N_{\text{trig}}}\,C_{\text{inf}}(\Delta x)\; \text{.}
\label{Method3A}
\end{eqnarray}
This is the same result as Eq.~(\ref{Method3}). 

To recap, we can avoid finite-acceptance effects by manipulating trigger particle acceptance and $\Delta x$ regions, or correct for the finite-acceptance effects in $(x_{\text{t}}, \Delta x)$ space by selectively copying the signal shape within certain region before averaging over $x_{\text{t}}$, assuming translational invariance of the signal. Although there is no general formula which relates the measured per-trigger yield and the per-trigger yield without finite-acceptance effects directly in $\Delta x$ space, we have found two exact formulas under specific conditions. If these methods in $\Delta x$ space are applied to the real analysis, we first generate the per-trigger yield, $C(\Delta\varphi, \Delta\eta)$, with certain trigger and associated particle conditions from every event, not concerning finite-acceptance effects. For the near-side structure, (1) if trigger particle distribution is constant, we can divide $C(\Delta\varphi, \Delta\eta)$ by $\frac{(A_{\text{a}}\star A_{\text{t}})(\Delta x)}{\Delta_{\text{t}}}$, or (2) if the trigger particle distribution is $\delta$-function-like, such as jet-hadron correlations, by $\frac{(A_{\text{a}}\star n_{\text{trig}}A_{\text{t}})(\Delta x)}{N_{\text{trig}}}$. Under the additional assumptions that $g(X,Y)$ depends only on $X-Y$ or $g(X,Y)$ is decomposed into $h(X)F(X-Y)$, the same methods can be applied for the away-side correlations. These new methods in $\Delta x$ space can be regarded as approximate formulas for more general cases and the validity depends on how close the correlated signal is to the assumed conditions, including translational invariance assumption. 

Coming back to $C_{\text{R}}$ and $C_{\text{trig,R}}$ from Eq.~(\ref{CrEq}) and (\ref{CtrigR}), we know that they are only different by a normalization factor and addition of a constant. What is intended from the ratio, $\frac{S(\Delta\varphi, \Delta\eta)}{B(\Delta\varphi, \Delta\eta)}$, used in both formulations is that it corresponds to $\frac{\rho_{\text{a,t}}(\varphi_{\text{t}}-\varphi_{\text{a}}; \eta_{\text{t}}- \eta_{\text{a}})}{\rho_{\text{a}}(\varphi_{\text{a}}, \eta_{\text{a}})\,\rho_{\text{t}}(\varphi_{\text{t}}, \eta_{\text{t}})}$. However, $S(\Delta\varphi, \Delta\eta)$ and $B(\Delta\varphi, \Delta\eta)$ are equivalent to $\rho_{\text{a,t}}(\varphi_{\text{t}}-\varphi_{\text{a}}; \eta_{\text{t}}- \eta_{\text{a}})$ and $\rho_{\text{a}}(\varphi_{\text{a}}, \eta_{\text{a}})\,\rho_{\text{t}}(\varphi_{\text{t}}, \eta_{\text{t}})$ with finite-acceptance effects, respectively. If finite-acceptance effects are factorized by the same function from $S(\Delta\varphi, \Delta\eta)$ and $B(\Delta\varphi, \Delta\eta)$, we can get $C_{\text{R}}$ correctly since finite-acceptance effects are canceled in the numerator and denominator. But these factorizability and cancellation in numerator and denominator do not hold in general cases. As we have shown, finite-acceptance effects depend on signal types and they may be non-factorizable. Thus, $C_{\text{R}}$ is merely an approximation of the intended correlation function. $C_{\text{trig,R}}$ should also be distinguished from its intended meaning, per-trigger yield. If $S(\Delta\varphi, \Delta\eta)$ in Eq.~(\ref{CtrigR}) consists of correlated signal and uncorrelated background, dividing by the mixed-event function may get rid of the uncorrelated background shape, but simultaneously distorts the correlated signal shape. 

\subsection{Discussion on true per-trigger yield}
\label{sec:TargetFunc}
So far, the derivations of the new methods are based on a few assumptions, such as translational and azimuthal invariance of the correlated signal, and the comparison with $C_{\text{inf}}$. In other words, $C_{\text{inf}}$ is considered as the true per-trigger yield, which is intended to be recovered by the finite-acceptance correction of the measured per-trigger yield.
Although the assumptions used for the derivations may not be satisfied in reality, the advantage of our formalism is to ensure mathematical completeness. The corrected result by design is free from finite-acceptance effects, but at the same time we need information outside of actual acceptance for the measurement. 
Thus, the validity of considering $C_{\text{inf}}(\Delta x)$ as a true per-trigger yield is closely related to the validity of translational invariance assumption. 
If translational invariance is assumed to hold only for the trigger acceptance range, a per-trigger yield $C_{\text{fin.trig}}$, which is defined with finite acceptance for the trigger particles and infinite acceptance for the associated particles, 
can be considered, especially in the case of near-side jet-like correlations. 
Mathematically, $C_{\text{fin.trig,single}}$ and $C_{\text{fin.trig}}$ can be written in a similar way to Eq.~(\ref{CidealSingle}) and Eq.~(\ref{Cideal}), but there is no exact formula to relate these two. Even though associated particle acceptance is infinite, there are still finite-acceptance effects in $C_{\text{fin.trig}}$. This is obvious when part of trigger particle distribution is within acceptance and others are not. However, $C_{\text{fin.trig}}$ will be considered in the Monte Carlo comparisons in the following section. 

\section{Test of alternative methods with Monte Carlo simulations}
\label{Sec:MC}
In the present section, we will apply the derived methods in $\Delta x$ space to Monte Carlo simulations and compare the results with $C_{\text{fin.trig}}$, per-trigger yield evaluated with finite acceptance for the trigger particles and infinite acceptance for the associated particles.
For convenience, we denote the correction method using
\begin{itemize}
\item the standard mixed-event technique~(Eq.~(\ref{CtrigR})) as Method~1,
\item the constant-signal ansatz with $\frac{(A_{\text{a}}\star A_{\text{t}})(\Delta x)}{\Delta_{\text{t}}}$~(Eq.~(\ref{Method2}) and (\ref{Method2A})) as Method~2,
\item and the $\delta$-function ansatz with $\frac{(A_{\text{a}}\star n_{\text{trig}}A_{\text{t}})(\Delta x)}{N_{\text{trig}}}$~(Eq.~(\ref{Method3}) and Eq.~(\ref{Method3A})) as Method~3
\end{itemize}
throughout the section.
Two sets of simulated data are analyzed: A simulation of di-jet events using PYTHIA~\cite{Sjostrand:2001yu} to test the methods on jet-like correlations (section~\ref{sec:MCPYTHIA}) and a simple MC toy model that creates a global correlation of all particles to an event plane to test the methods on flow-like correlations (section~\ref{sec:MCColl}). 

\begin{table}[b!f] \centering
 \begin{tabular}{|c|}
   \hline
          PYTHIA (ver. 6.205) Di-jet process\\
   \hline
     Proton beam energy =  2.76 TeV \\
     $10^7$ events \\
     No initial gluon radiation \\
     Intrinsic $k_{\text{T}} = 0$ \\
     Minimum jet $p_{\text{T}} = 10$ GeV/$c$ \\
     No vertex smearing \\
     Structure function $=$ CTEQ4L \\
     \hline
 \end{tabular}
 \caption{\label{tab:Pythia} Settings used for the PYTHIA generator-level simulation.}
\end{table}

\subsection{PYTHIA simulation}
\label{sec:MCPYTHIA}

To test the applicability of the correction methods in the case of jet-like correlations, PYTHIA di-jet events are generated which contain back-to-back jets with the same $p_{\text{T}}$.
Details of the PYTHIA Monte Carlo configurations can be found in Table~\ref{tab:Pythia}. To have clean back-to-back jet correlation signal, each event is set to have two jets with jet $p_{\text{T}}$, defined as $p_{\text{T}}$ of the hard scattered parton at the origin of the jet, larger than 10 GeV/$c$ and intrinsic transverse momentum ($k_{\text{T}}$) equal 0. We have assumed two different $\eta$-acceptances, $[-2, 2]$ and $[0,4]$, to test the new methods and estimated $C_{\text{inf}}$ and $C_{\text{fin.trig}}$, not applying any $\eta$-cut and applying $\eta$-cut to the trigger particles, respectively. 

\begin{figure}[t]
\centering
\includegraphics[width=0.49\linewidth]{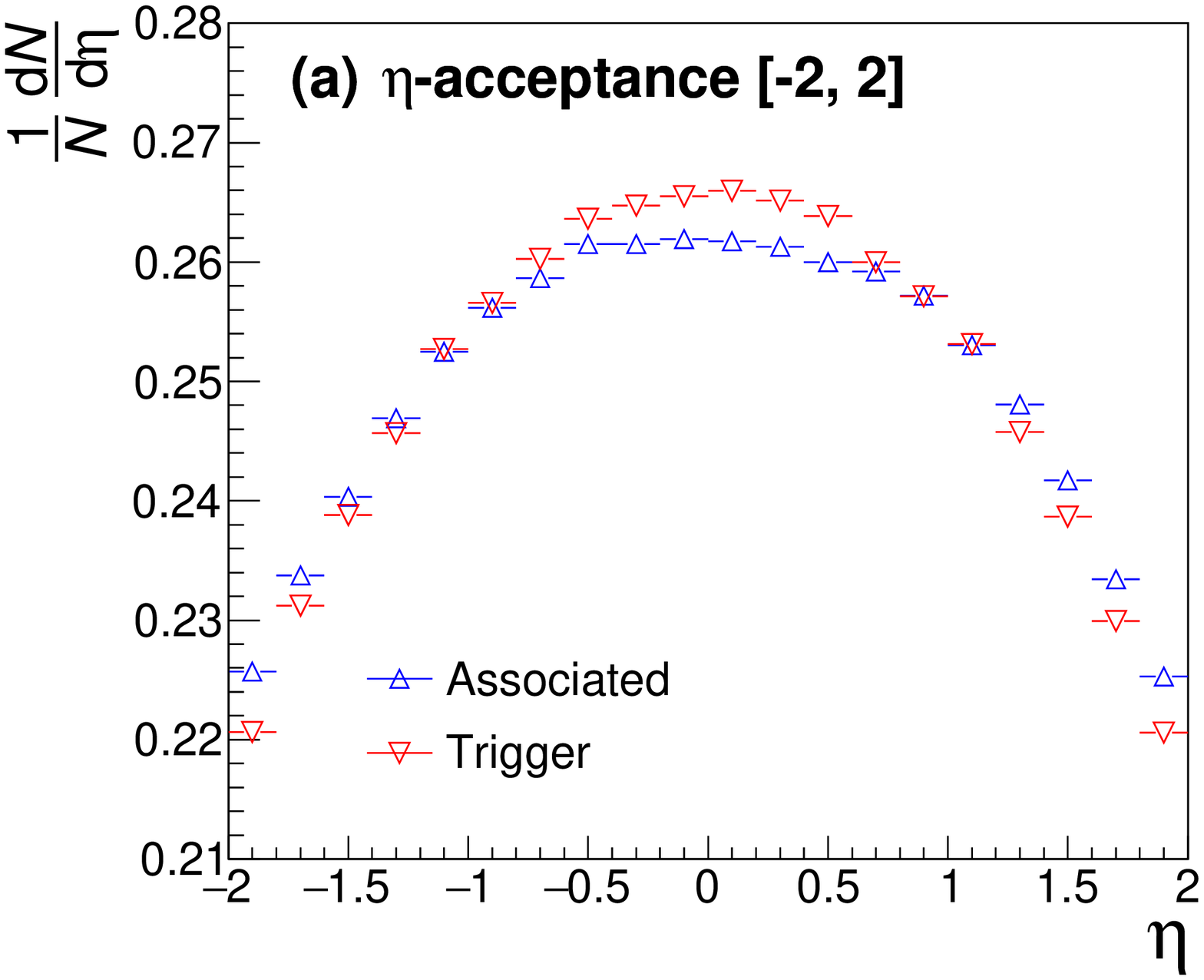}
\includegraphics[width=0.49\linewidth]{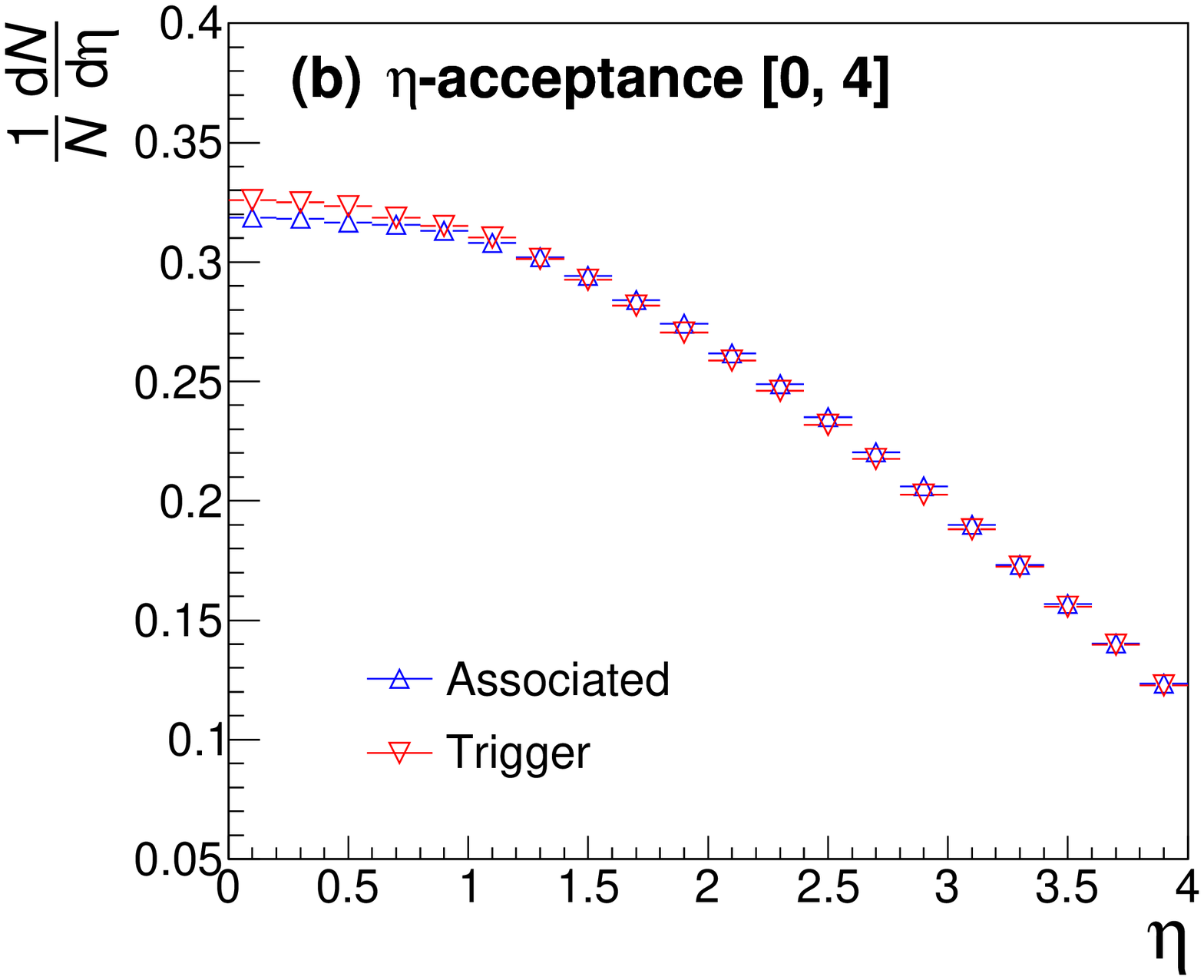}
\caption{\label{fig:pythia_eta} $\eta$-distribution of associated particles ($1.0 \text{ GeV}/c < p_{\text{T, assoc.}} < 2.0 \text{ GeV}/c$) and trigger particles ($2.0 \text{ GeV}/c < p_{\text{T, trig}} < 50.0 \text{ GeV}/c$) with the (a)~$\eta$-acceptance $[-2,2]$ and (b)~$\eta$-acceptance $[0, 4]$.}
\end{figure}

There are many possible choices for trigger and associated particle conditions, such as $p_{\text{T}}$ and particle species. For our examples, we choose to use every final-state particle with $2.0 \text{ GeV}/c < p_{\text{T, trig}} < 50.0 \text{ GeV}/c$ and $1.0 \text{ GeV}/c < p_{\text{T, assoc.}} < 2.0 \text{ GeV}/c$ for the current analysis. This means that the Method~3 only holds approximately because low-$p_{\text{T}}$ trigger particles are not aligned with the jet axis and the trigger particle $\eta$ distributions might contain a contribution from soft particle production. Figure~\ref{fig:pythia_eta} shows normalized $\eta$-distributions of trigger and associated particles for two $\eta$-acceptances, and we can see that they are not uniform within the acceptances. Trigger (associated) particle distribution is normalized by the number of trigger (associated) particles. Hence, also Method~2 can only be considered an approximation, since it is derived for the case of a uniform signal distribution.

\begin{figure}[t]
\centering
\includegraphics[width=0.49\linewidth]{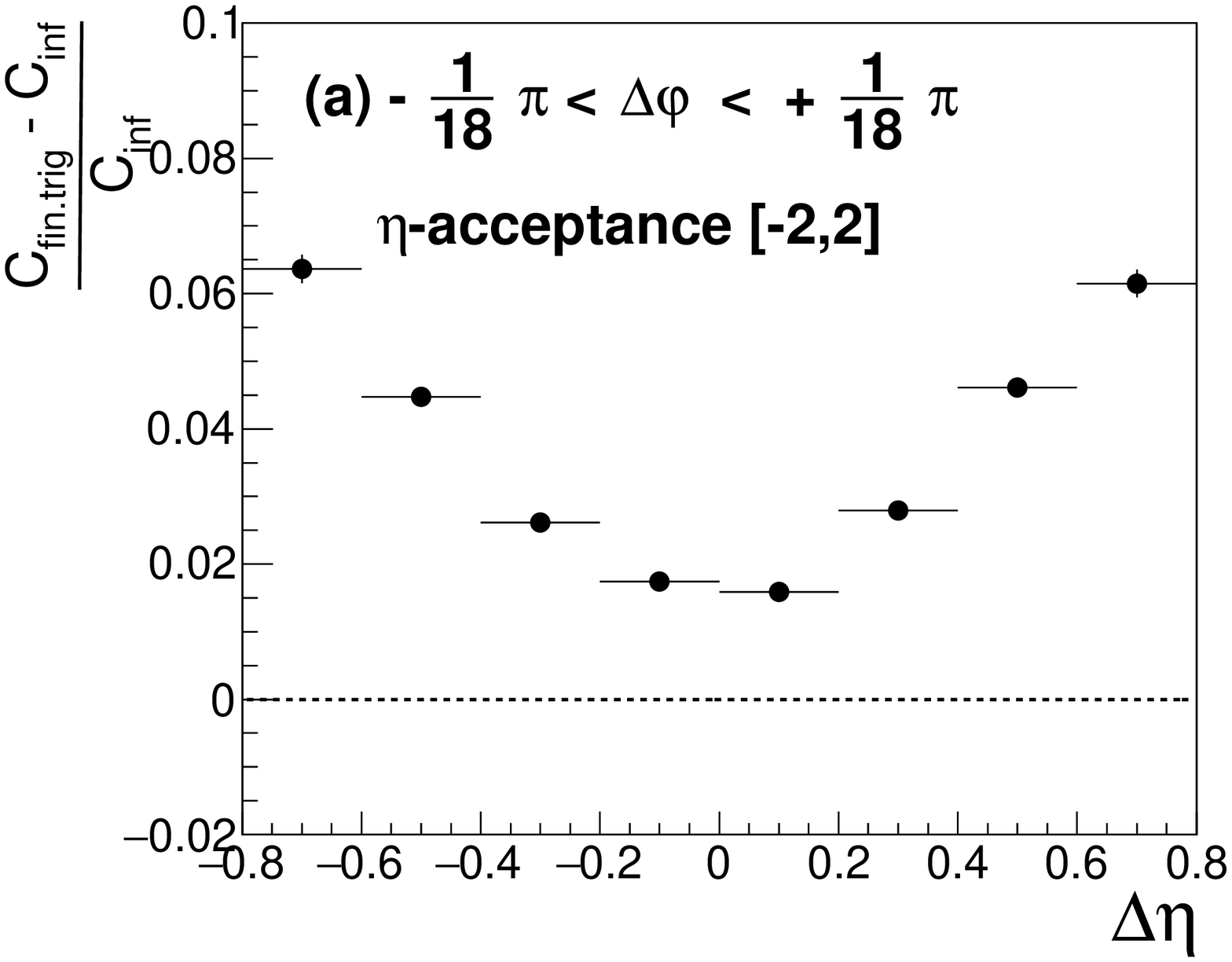}
\includegraphics[width=0.49\linewidth]{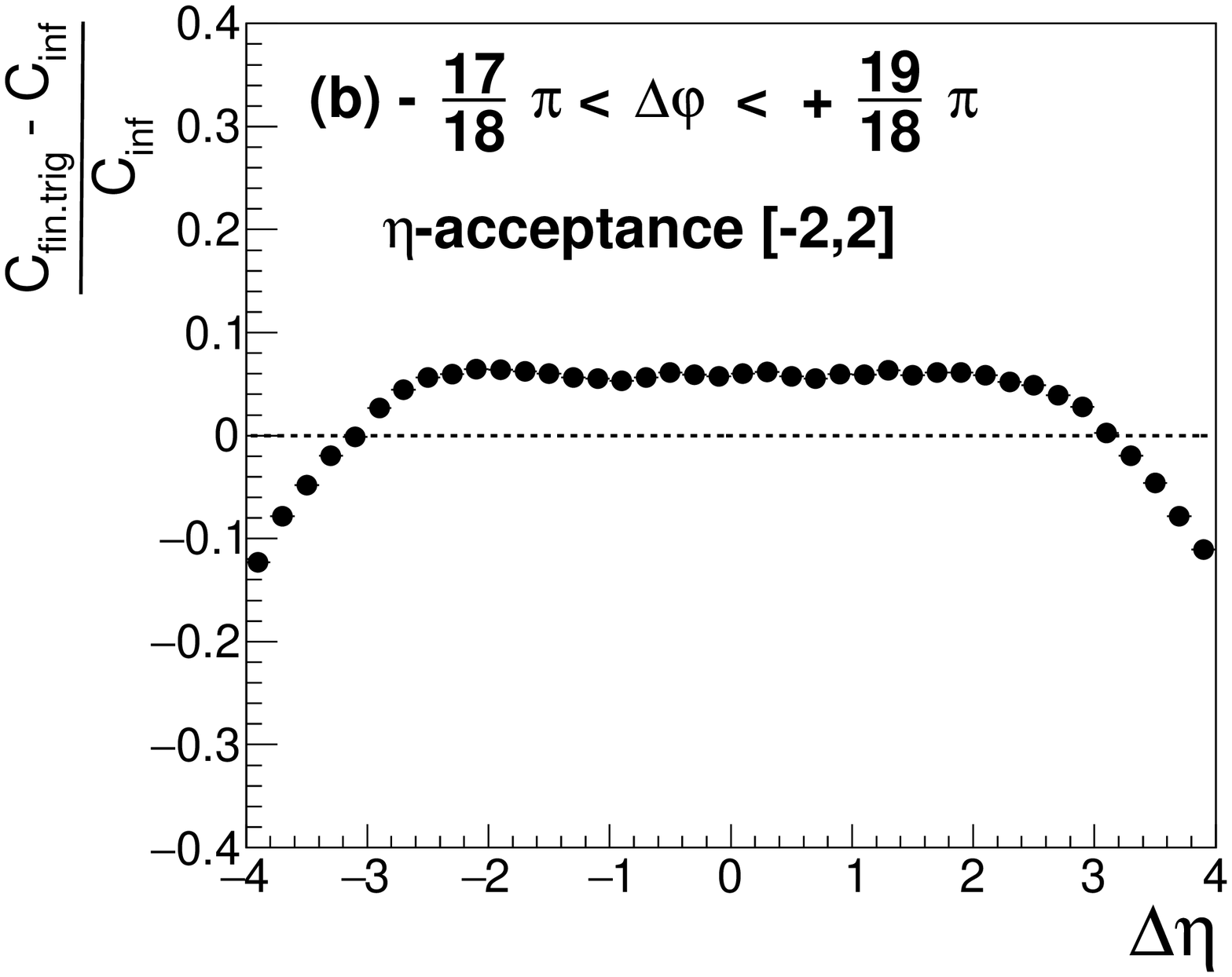}
\includegraphics[width=0.49\linewidth]{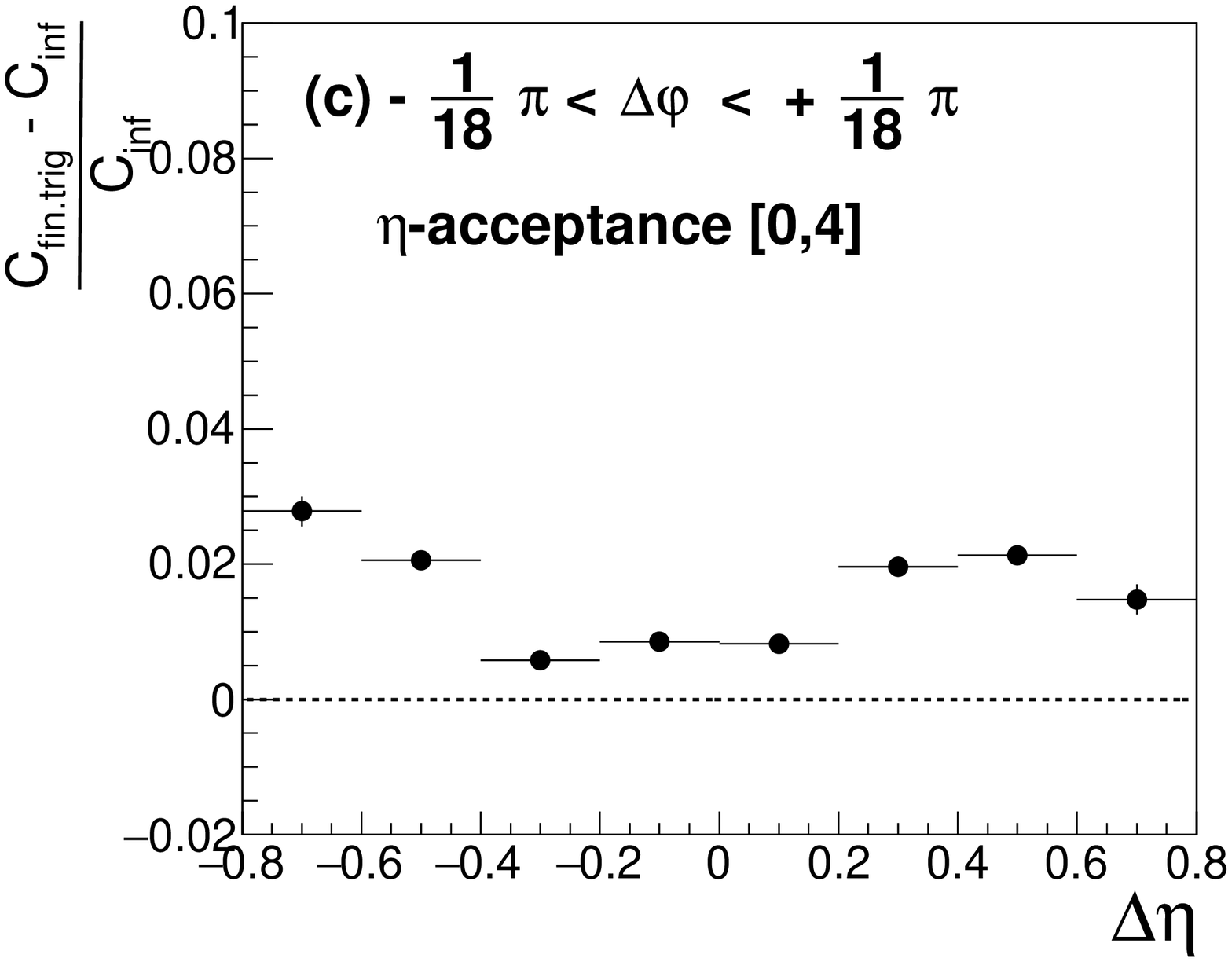}
\includegraphics[width=0.49\linewidth]{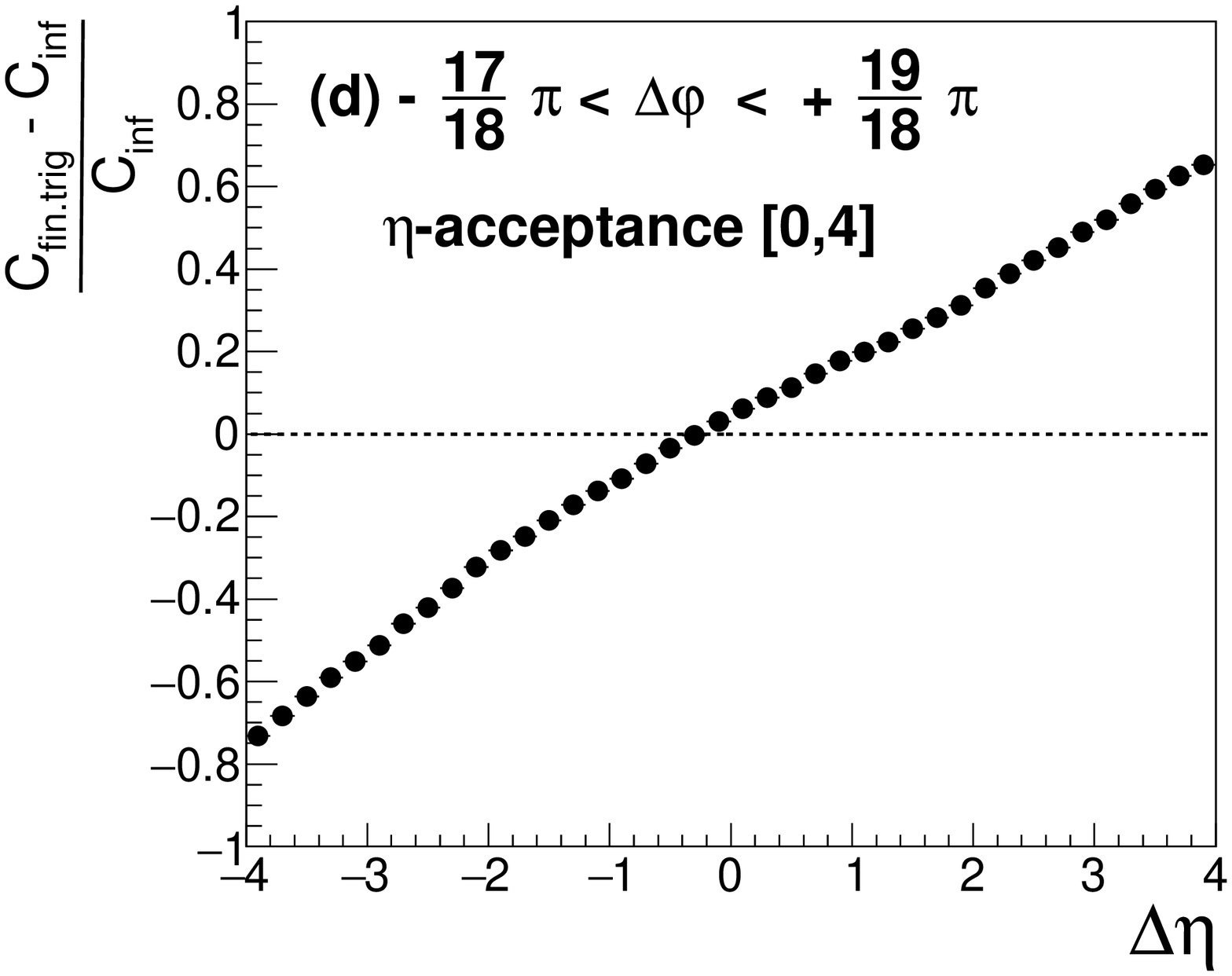}
\caption{\label{fig:pythia_finvsinf} $\Delta\eta$ projections of $(C_{\text{fin.trig}} - C_{\text{inf}})/C_{\text{inf}}$ on (a)~near and (b)~away side with $\eta$-acceptance $[-2, 2]$ and (c)~near and (d)~away side with $\eta$-acceptance $[0,4]$.}
\end{figure}

As described in section~\ref{sec:TargetFunc}, $C_{\text{inf}}$ and $C_{\text{fin.trig}}$ are different especially when translational invariance assumption is not fully satisfied. 
Figure~\ref{fig:pythia_finvsinf} shows comparisons between $C_{\text{inf}}$ and $C_{\text{fin.trig}}$ with $\Delta\eta$-projections of 
\begin{equation}
\frac{C_{\text{fin.trig}} - C_{\text{inf}}}{C_{\text{inf}} }
\end{equation}
for near and away side for two acceptances. 
For the near side the $-\pi/18 < \Delta\varphi < \pi/18$ region, and for the away side the $(1-1/18)\pi < \Delta\varphi < (1+1/18)\pi$ region is projected onto the $\Delta\eta$-axis. 
In PYTHIA events, jets located at large $|\eta|$ are expected to have narrower fragmented particle distribution. 
$C_{\text{inf}}$ includes these jets at large $|\eta|$ in the estimation, while $C_{\text{fin.trig}}$ does not. 
This results in larger value of $C_{\text{fin.trig}}$ than $C_{\text{inf}}$ on the near side especially at large $|\Delta\eta|$. 
In the case of  $[0,4]$ $\eta$-acceptance, away-side structure of $C_{\text{fin.trig}}$ is expected to be asymmetric with respect to $\Delta\eta=0$, while $C_{\text{inf}}$ is symmetric. 
This intrinsic difference by definition is shown in Fig.~\ref{fig:pythia_finvsinf}~(d). 
In the following, corrected per-trigger yields will be compared to $C_{\text{fin.trig}}$ as the difference between $C_{\text{inf}}$ and $C_{\text{fin.trig}}$ is understood. 

\begin{figure}
\centering
\includegraphics[width=0.49\linewidth]{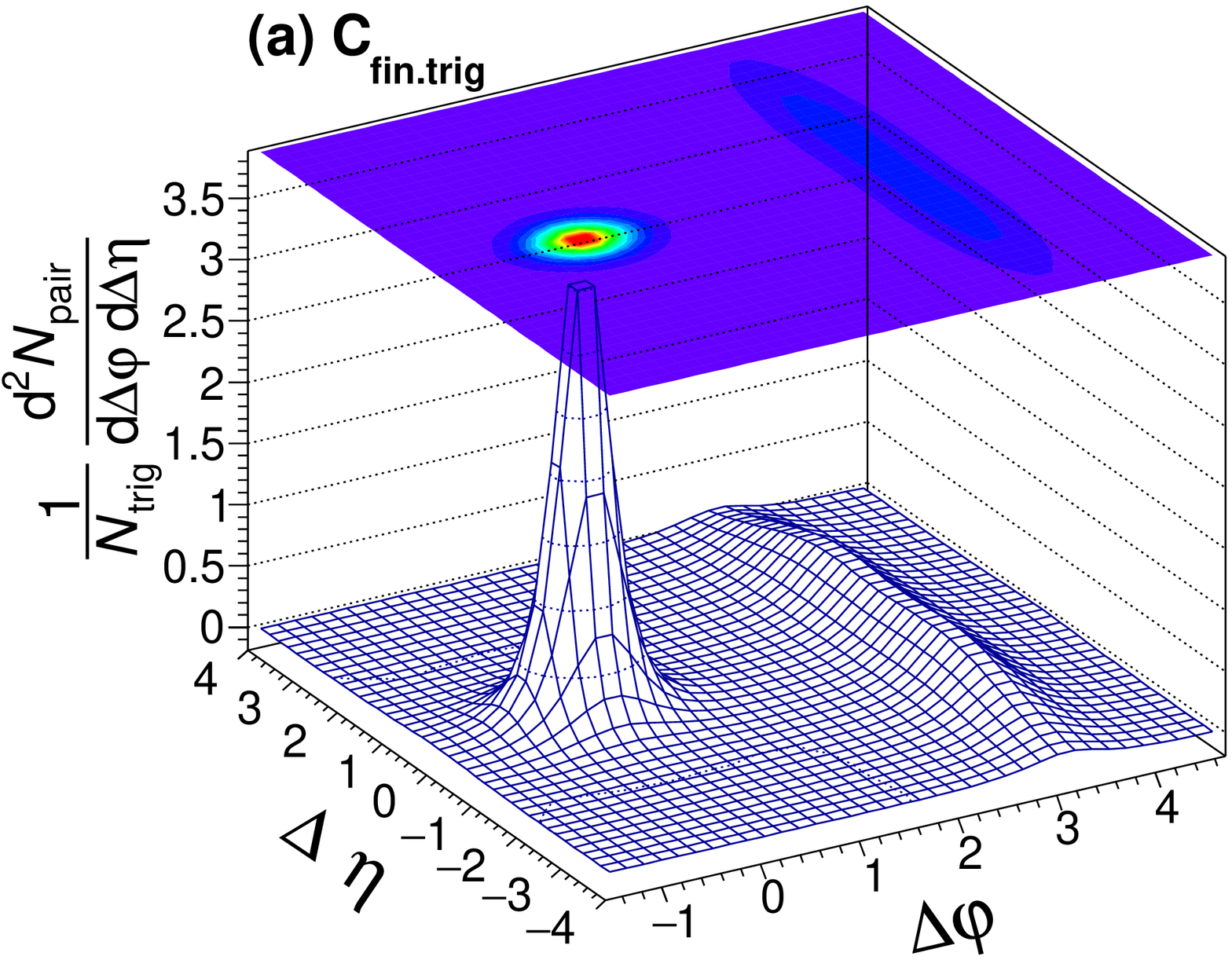}
\includegraphics[width=0.49\linewidth]{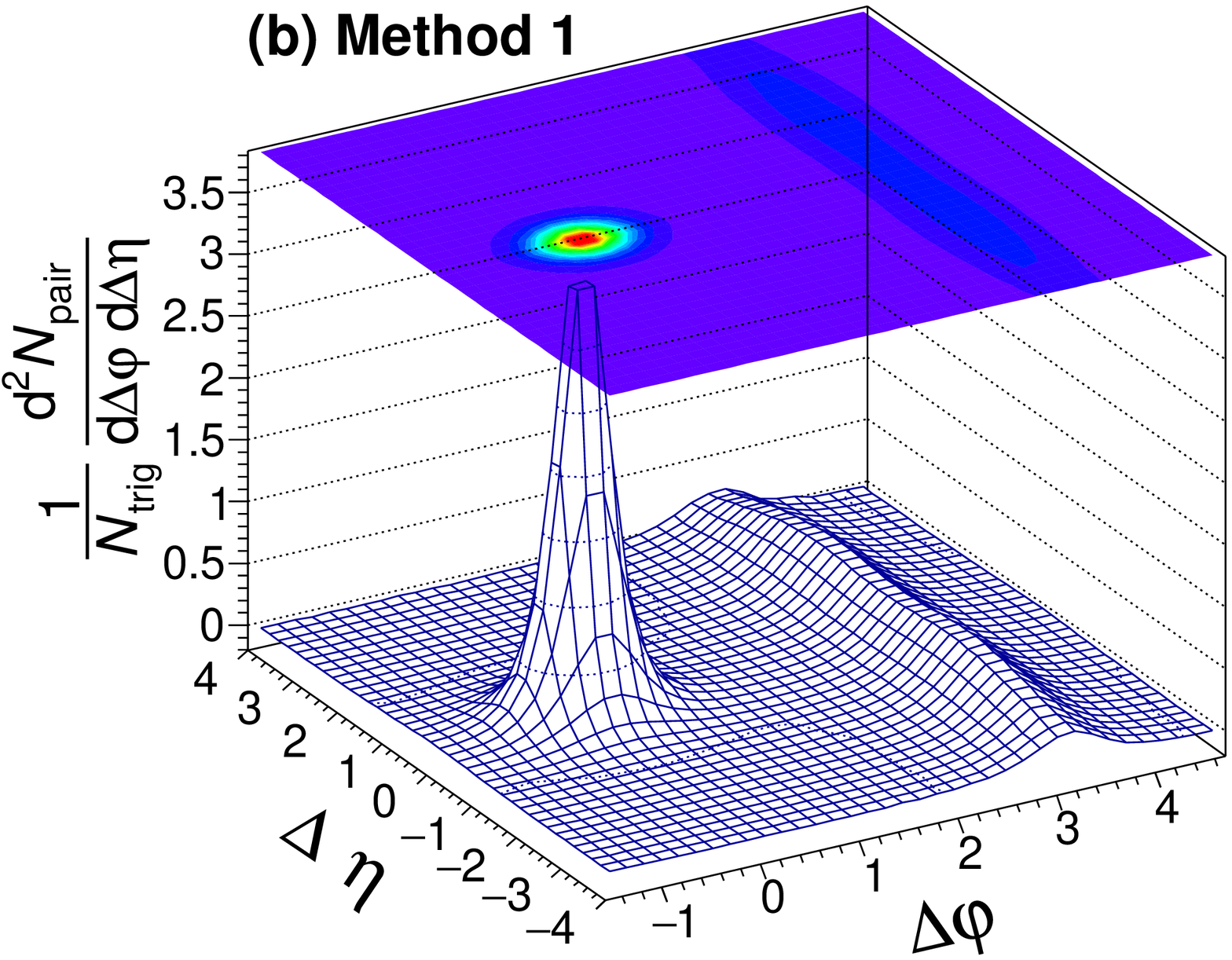}
\includegraphics[width=0.49\linewidth]{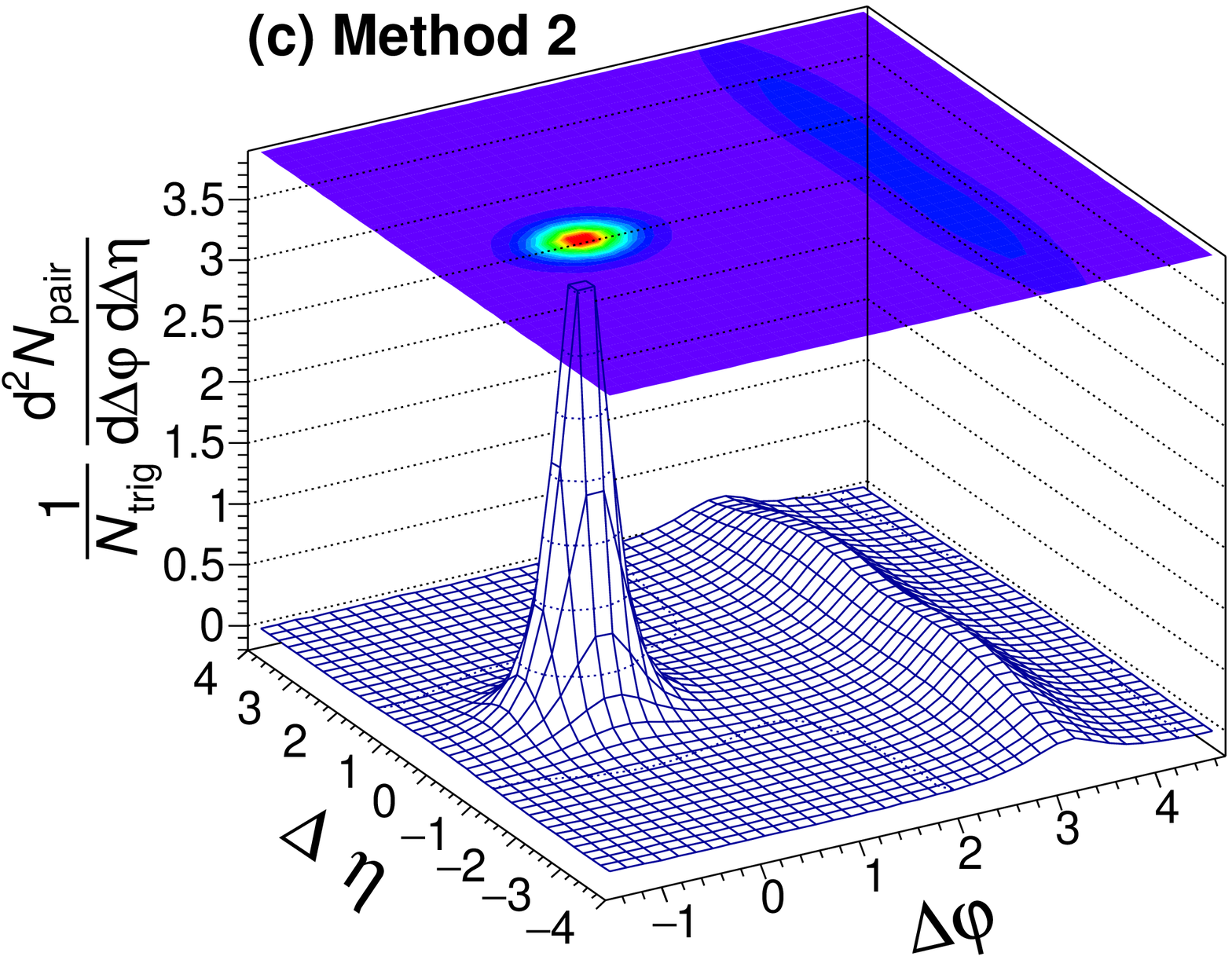}
\includegraphics[width=0.49\linewidth]{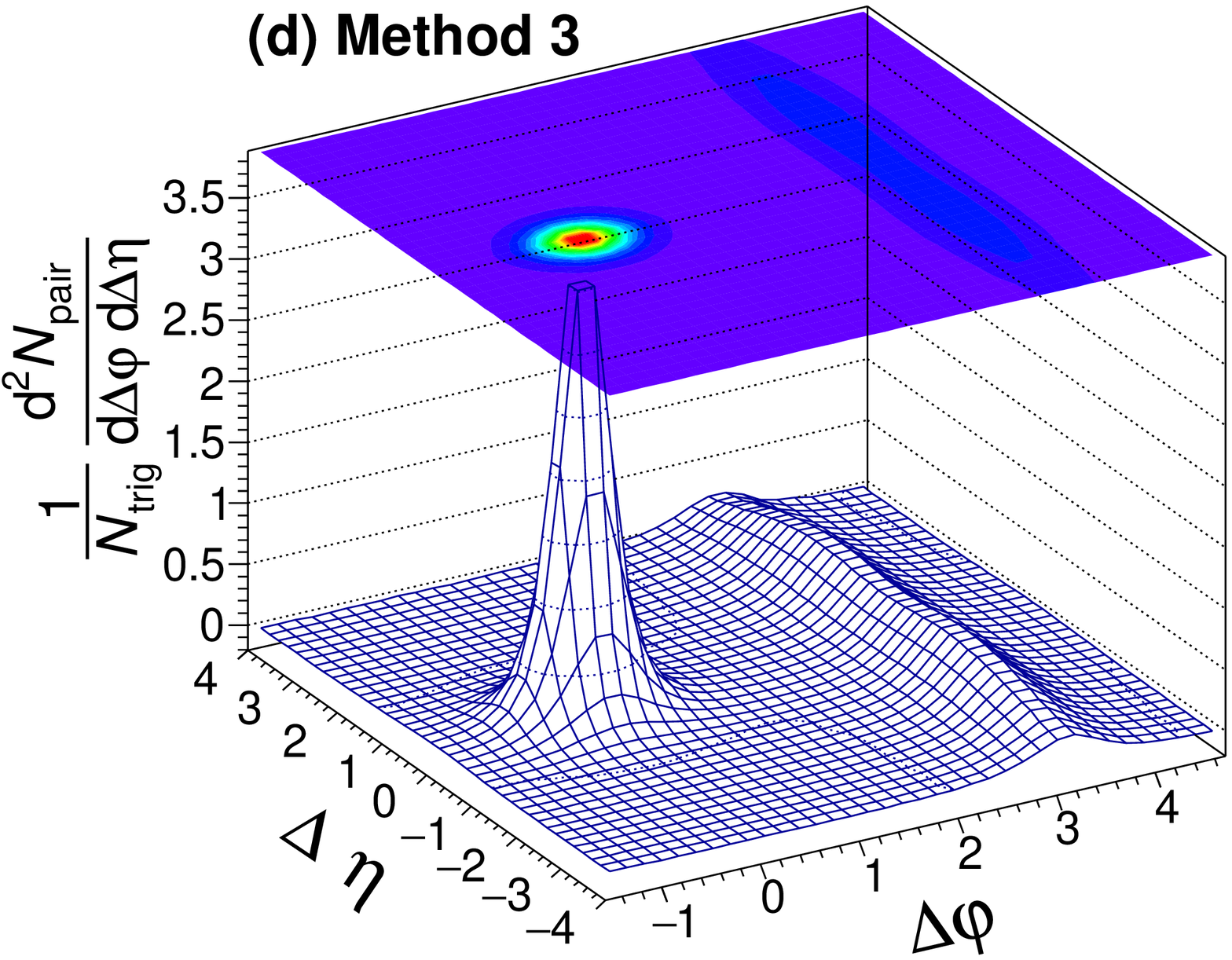}
\caption{\label{fig:pythia_22_2D} (a)~$C_{\text{fin.trig}}$; (b)~per-trigger yield from Method~1~(standard), (c)~Method~2~(uniform), and (d)~Method~3~($\delta$-function) for an $\eta$-acceptance of $[-2,2]$.}
\end{figure}

\begin{figure}
\centering
\includegraphics[width=0.49\linewidth]{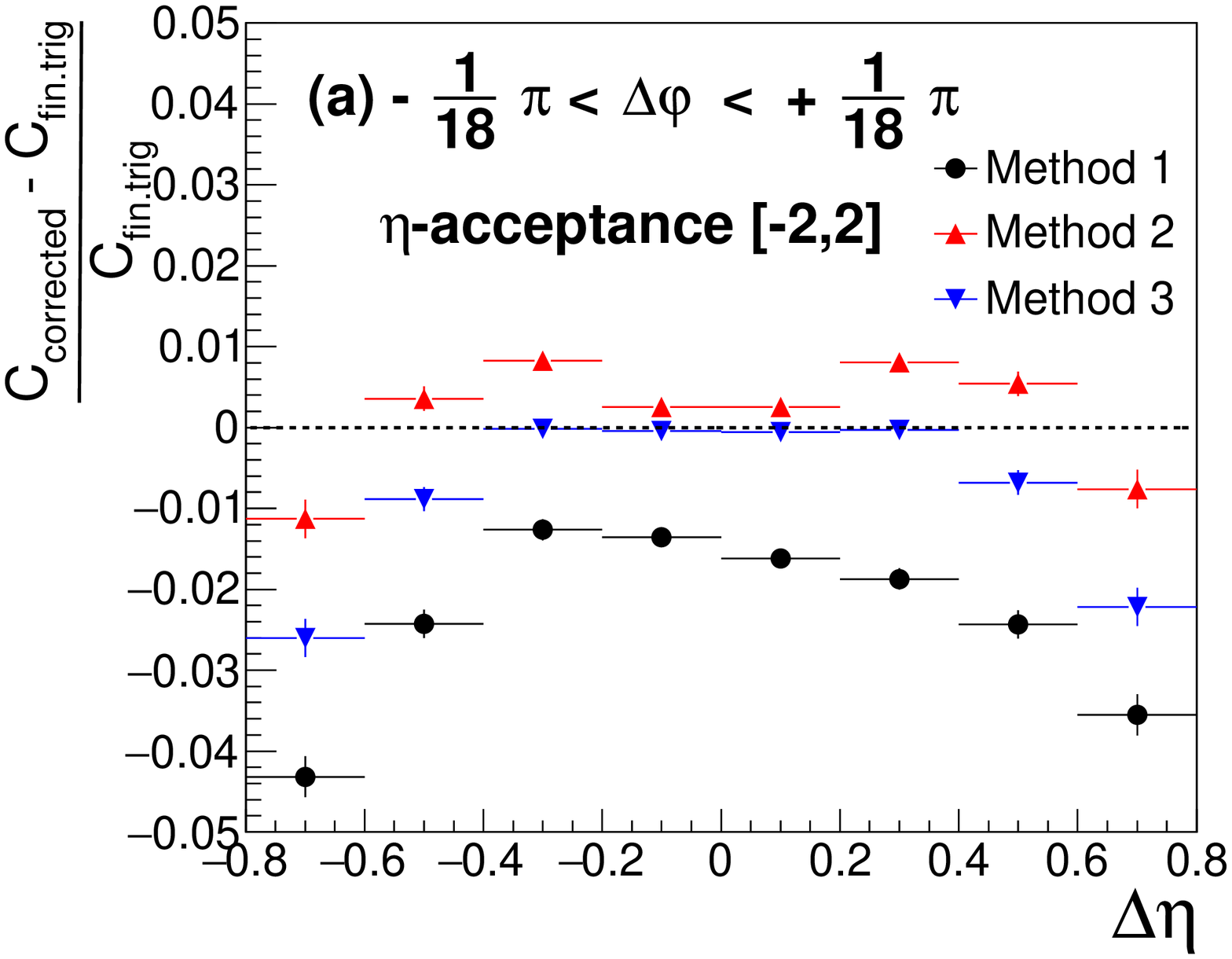}
\includegraphics[width=0.49\linewidth]{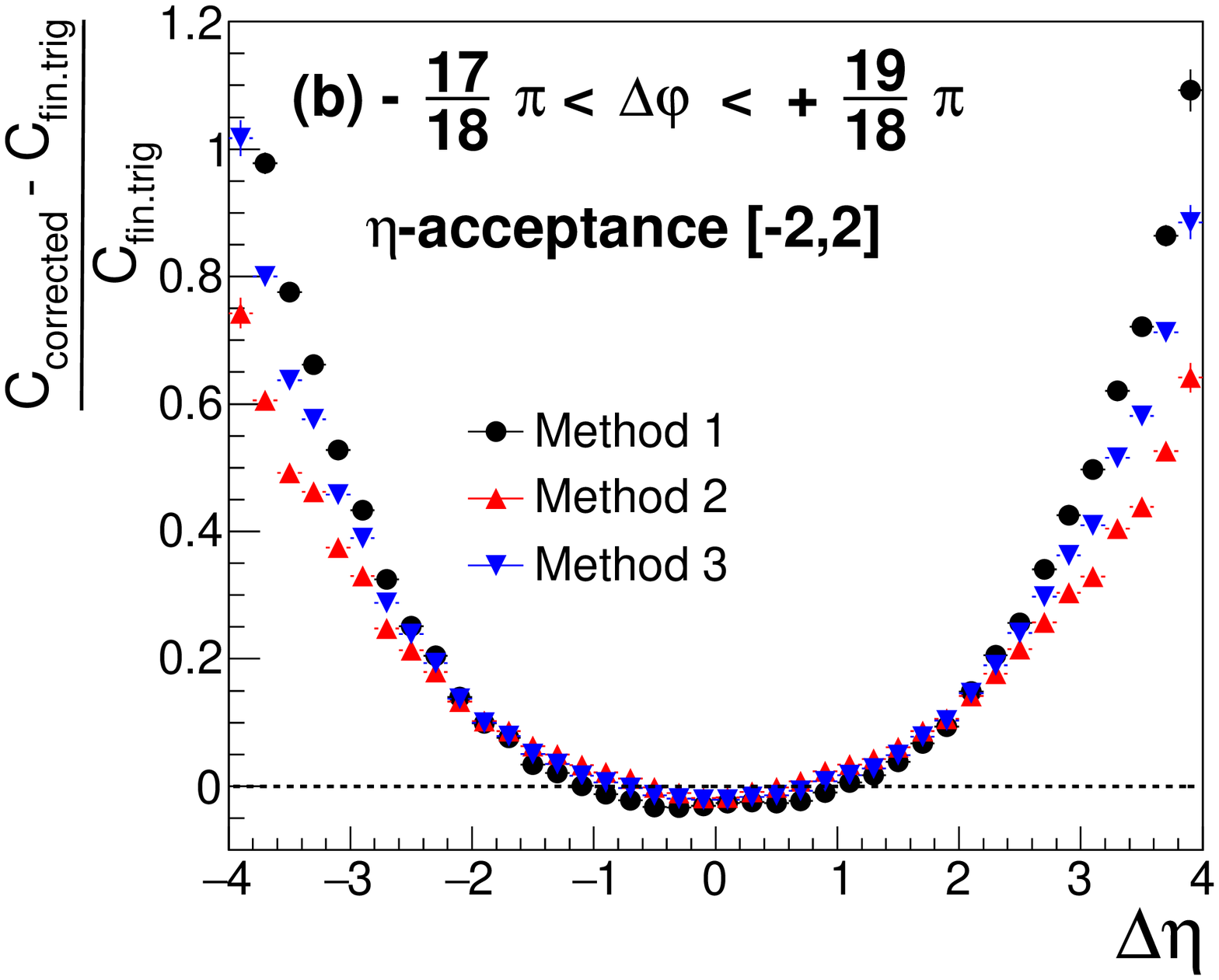}
\caption{\label{fig:pythia_22_Ratio} $\Delta\eta$ projections of $(C_{\text{corrected}} - C_{\text{fin.trig}})/C_{\text{fin.trig}}$ on (a)~near and (b)~away side for the comparison of the per-trigger yields presented in Figure~\ref{fig:pythia_22_2D}.}
\end{figure}

Figure~\ref{fig:pythia_22_2D} shows the $C_{\text{fin.trig}}$ and per-trigger yields from three correction methods after subtraction of a scaled mixed-event function as described before in Section~\ref{ch3-1} for an $\eta$ acceptance of $[-2,2]$.
Figure~\ref{fig:pythia_22_Ratio} shows the $\Delta\eta$ projections of
\begin{equation}
\frac{C_{\text{corrected}} - C_{\text{fin.trig}}}{C_{\text{fin.trig}} }
\end{equation}
for near and away side, while the same $\Delta\varphi$ regions as Fig.~\ref{fig:pythia_finvsinf} are used for $\Delta\eta$-projections. 
From Figure~\ref{fig:pythia_22_Ratio}, we conclude that results from three methods do not significantly deviate from $C_{\text{fin.trig}}$ on the near side, unlike on the away side, where for larger $\Delta\eta$ ranges significant deviations become apparent. 
In particular, the discrepancy on the away side is related to the violation of the initial assumptions, such as $g(X,Y)=G(X-Y)$ or $h(X)F(X-Y)$, and $\delta$-function trigger-particle distribution. 
However, we note that finite-acceptance corrections with Method~2 and Method~3 do not depend on the shape of mixed-event function, which attempts to describe the corresponding shape of the uncorrelated particle production.
It is generally expected that the corrections are more accurate at regions of smaller $|\Delta\eta|$ than the acceptance window, as less finite-acceptance effects are involved. 

\begin{figure}
\centering
\includegraphics[width=0.49\linewidth]{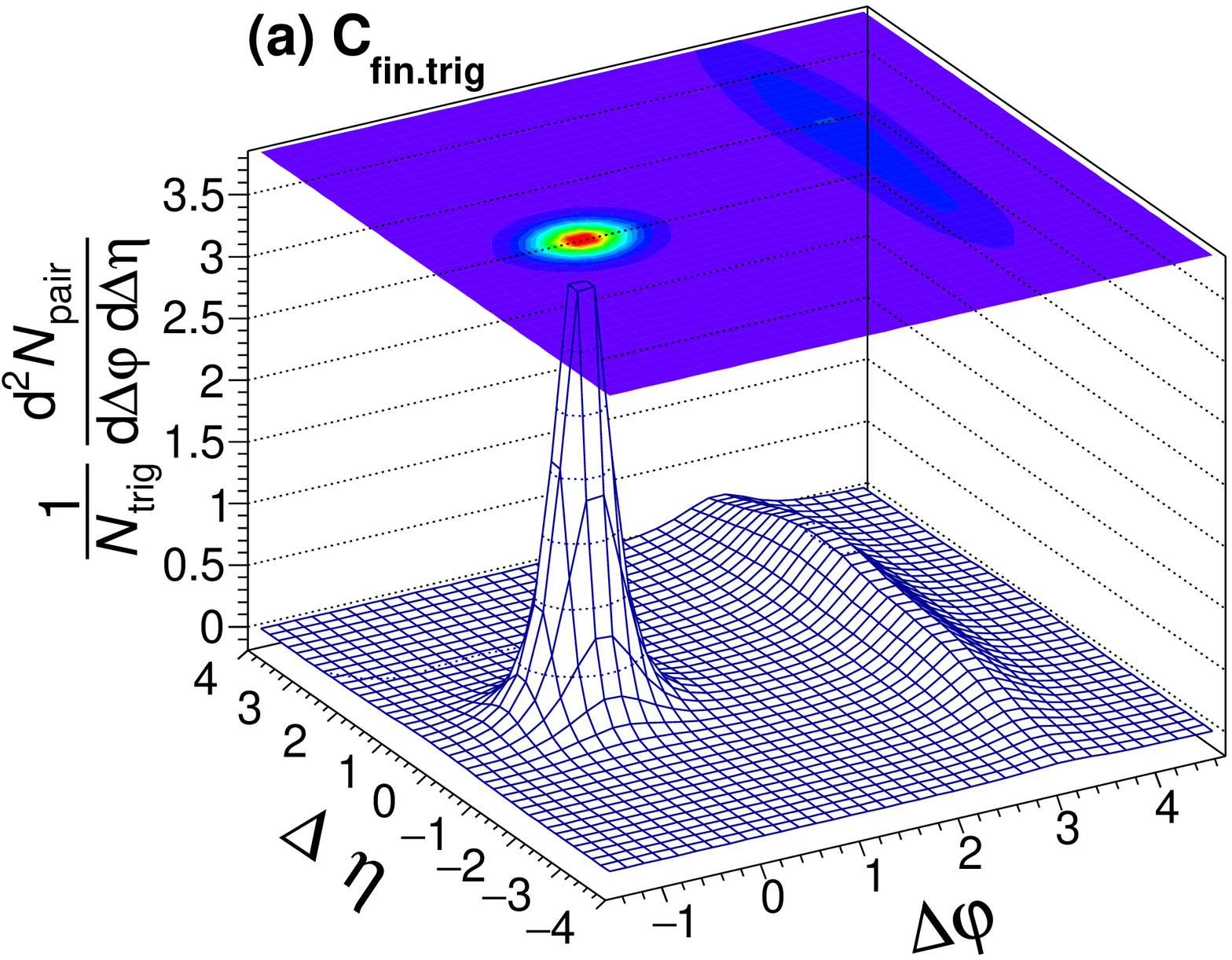}
\includegraphics[width=0.49\linewidth]{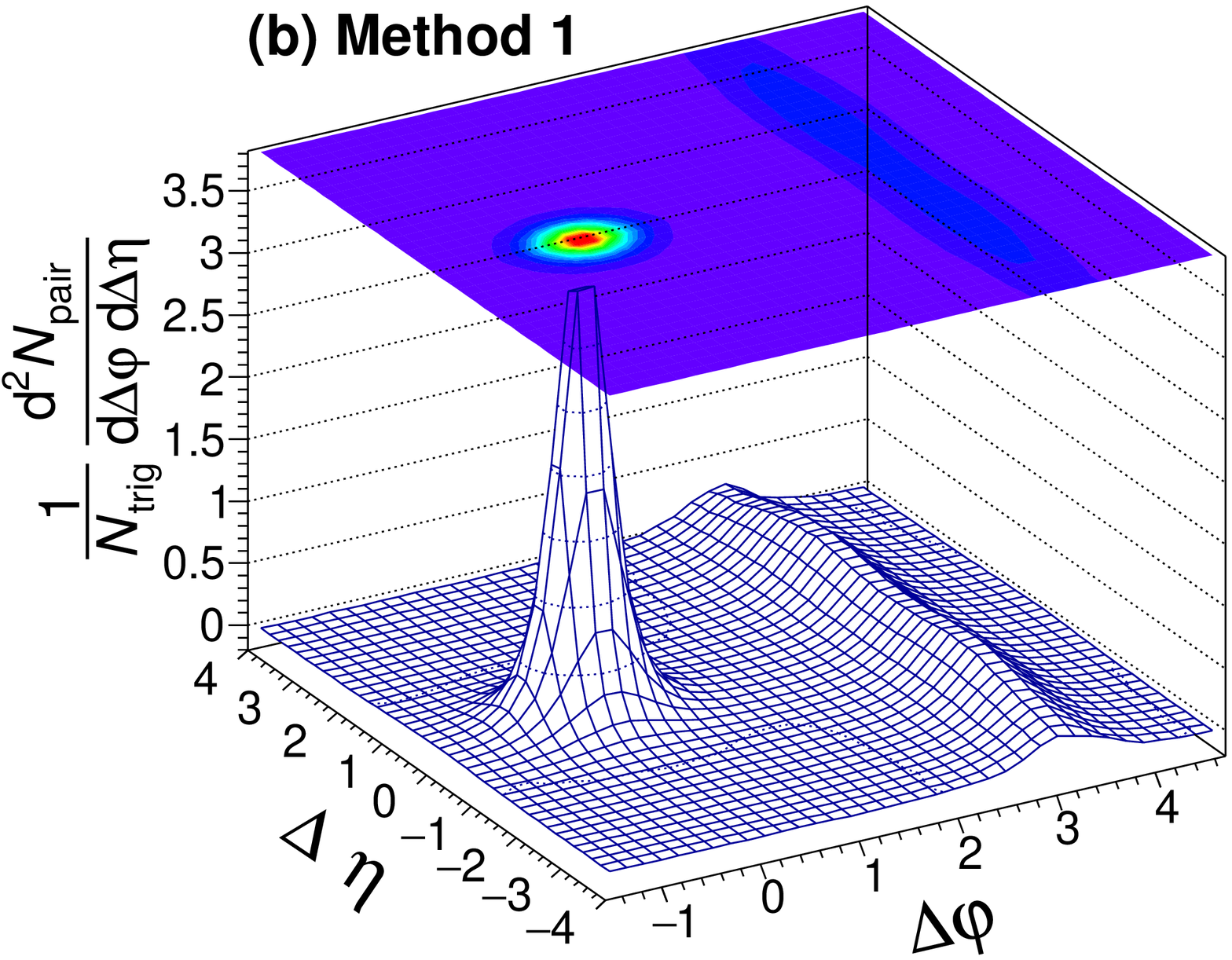}
\includegraphics[width=0.49\linewidth]{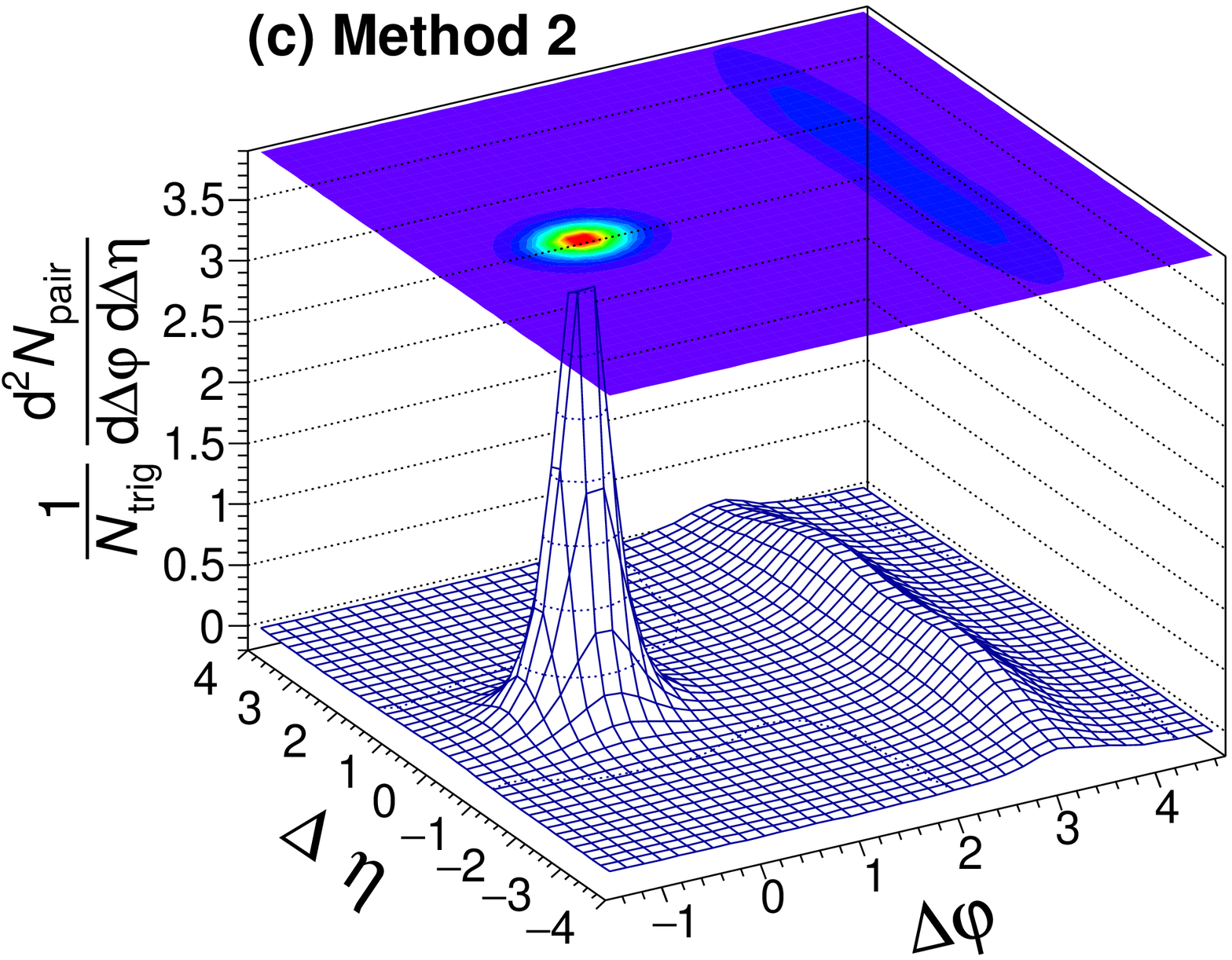}
\includegraphics[width=0.49\linewidth]{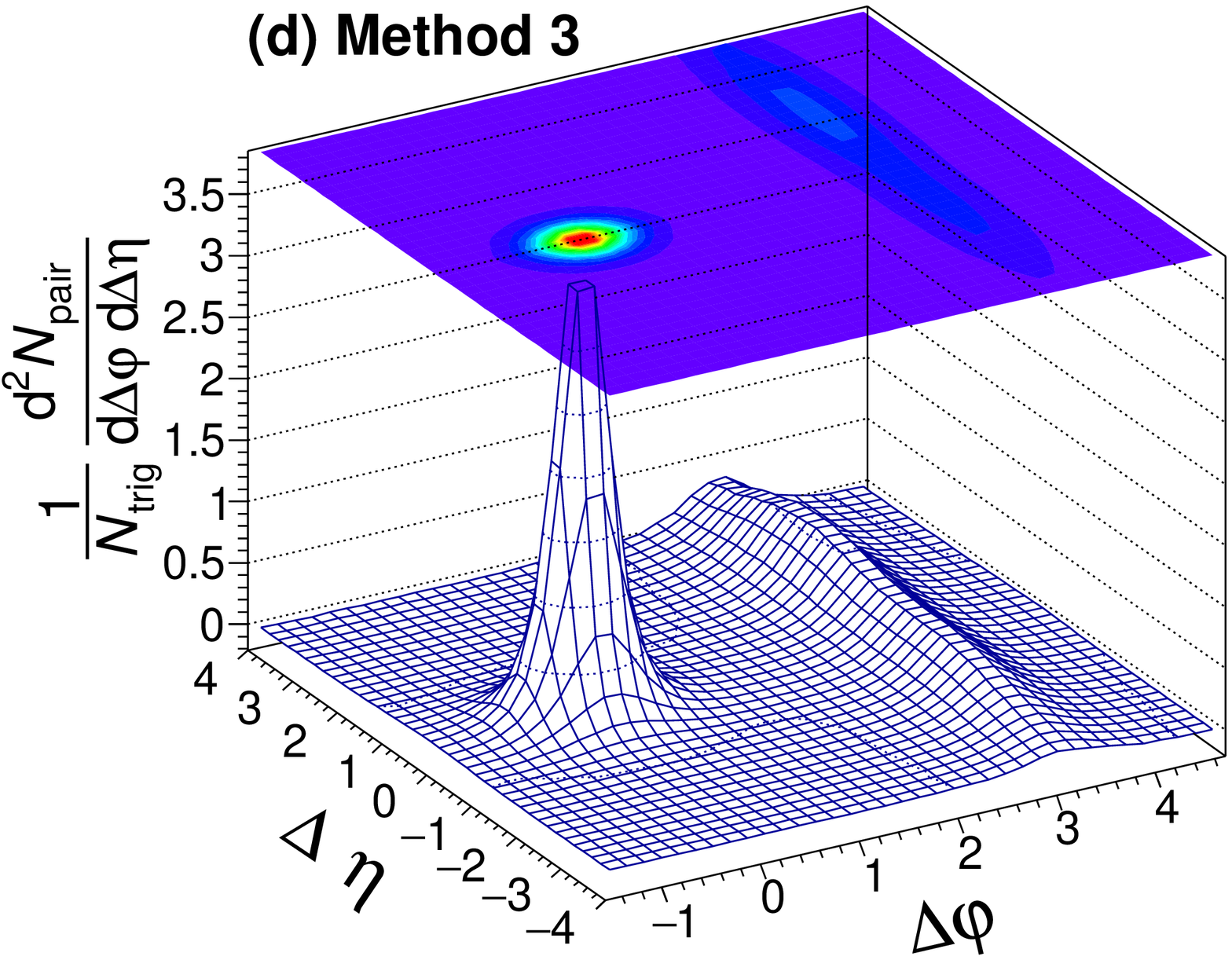}
\caption{\label{fig:pythia_04_2D} (a)~Per-trigger yield evaluated with infinite $\eta$ acceptance; (b)~per-trigger yield from Method~1~(standard), (c)~Method~2~(uniform), and (d)~Method~3~($\delta$-function) for an $\eta$-acceptance of $[0,4]$.}
\end{figure}

\begin{figure}
\centering
\includegraphics[width=0.49\linewidth]{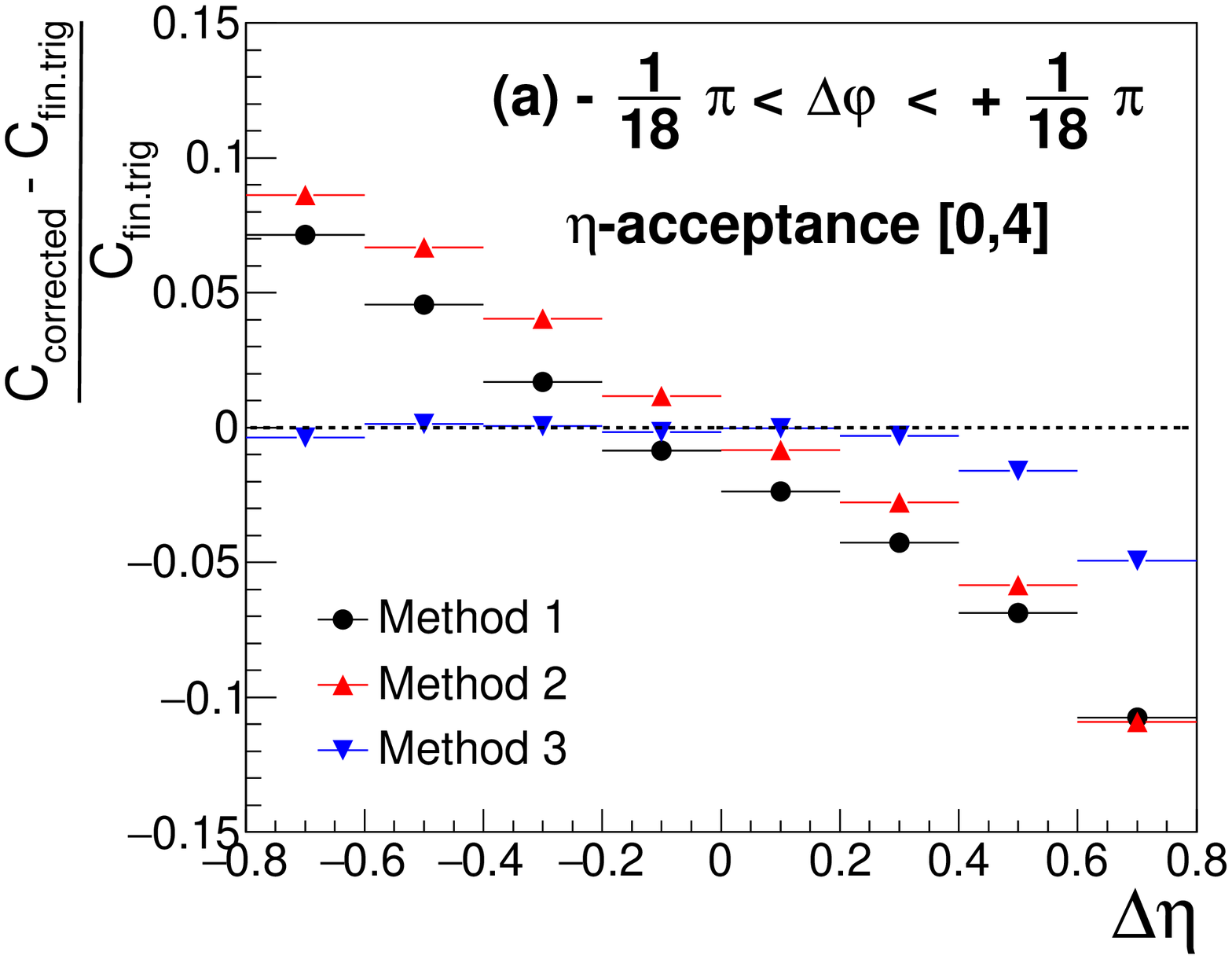}
\includegraphics[width=0.49\linewidth]{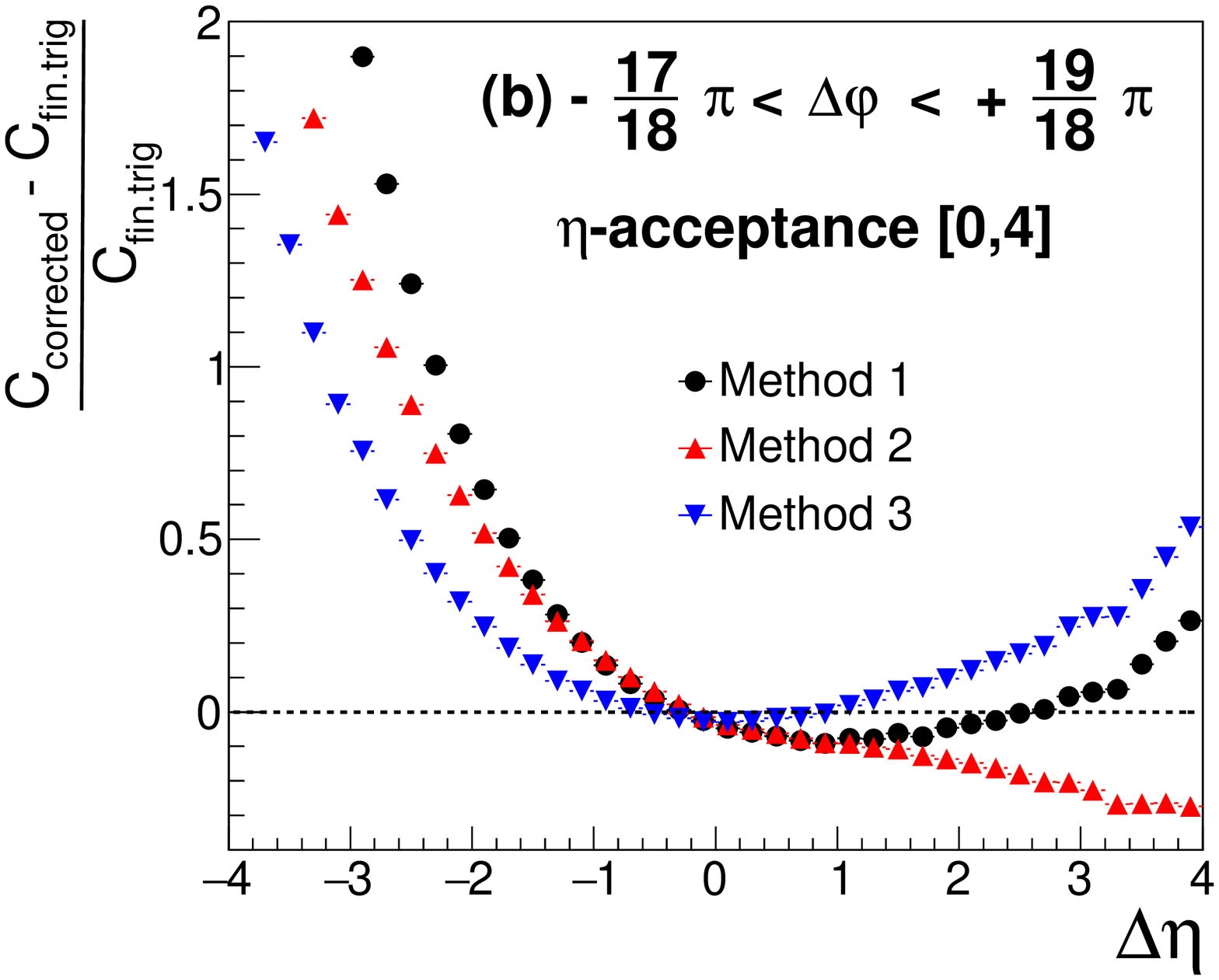}
\caption{\label{fig:pythia_04_Ratio} $\Delta\eta$ projections of (a)~near and (b)~away side for the comparison of the per-trigger yields presented in Figure~\ref{fig:pythia_04_2D}.}
\end{figure}

Figure~\ref{fig:pythia_04_2D} and Figure~\ref{fig:pythia_04_Ratio} correspond to Figure~\ref{fig:pythia_22_2D} and Figure~\ref{fig:pythia_22_Ratio}, respectively, but with $\eta$-acceptance $[0, 4]$ instead of $[-2, 2]$.
Particle distributions in $\eta$ in this case are asymmetric within the acceptance as shown in Figure~\ref{fig:pythia_eta}~(b), which is also the case for proton--nucleus collisions.
From Figure~\ref{fig:pythia_04_Ratio}, we observe that Method~3 roughly recovers the symmetry of near-side jet shape while the other two methods produce more asymmetric shapes. 
The asymmetry in shape is due to the asymmetric trigger distribution, and Method~3 is intended to reduce this effect. 
Asymmetric deviations are more obvious at large $\Delta\eta$. 

\subsection{Collective MC simulation}
\label{sec:MCColl}
In flow analyses, $v_2$ is an important observable and there are many approaches to evaluate $v_2$ from various collision systems~\cite{Ollitrault:1992bk,Poskanzer:1998yz}. To reduce non-flow effects, subtraction of the lowest multiplicity class from a higher multiplicity class is used under the assumption that per-trigger yields in the lowest multiplicity class are dominated by the non-flow signal and this signal is independent to the centrality class, or only large $\Delta\eta$ parts of per-trigger yields are considered where non-flow signal on the near-side cannot reach~\cite{ALICE:2PC:pPb}. However, due to the statistics, $v_2$ is commonly evaluated after per-trigger yields or correlation functions are projected into the $\Delta\varphi$ axis in a two-particle correlation analysis. In other words, after projecting the $C(\Delta\varphi, \Delta\eta)$ onto the $\Delta\varphi$ axis,  the azimuthal anisotropy harmonics are extracted from a Fourier decomposition,
\begin{equation}
\frac{1}{N_{\text{trig}}}\frac{\text{d}N_{\text{pair}}}{\text{d}\Delta\varphi}=\frac{N_{\text{assoc.}}}{2\pi}\left[1+\sum\limits_{n}2V_{n\Delta}\cos(n\Delta\varphi)\right] \;\text{,}
\end{equation}
and $v_2$ is calculated from $V_{2\Delta}$.

To evaluate the influence of finite-acceptance effects and the correction methods on the extracted $v_2$, a toy Monte Carlo model is used.
In every event, the $\eta$ distribution of particles is required to follow a common Gaussian function ($\mathrm{d}N/\mathrm{d}\eta\sim\text{exp}(-\eta^2/(2\sigma^2))$) with $\sigma=3$.
The $\varphi$ distribution is different from the common flow toy Monte Carlo simulation, which uses $\mathrm{d}N/\mathrm{d}\varphi\sim\big(1 + 2\,v_2\,\text{cos}(2(\varphi - \varphi_0))\big)$ with constant $v_2$.
To introduce a $\Delta\eta$ dependence in $v_2$, we randomly choose one particle in each event, and denote its $\eta$ value as $\eta_{\text{ref}}$. Then the $\varphi$ distribution of that event follows $\sim\big(1 + 2\,a_2(\eta)\,\text{cos}(2(\varphi - \varphi_0))\big)$, with
\begin{eqnarray}
 a_{2}(\eta) =
  \begin{cases}
     0.3\,\frac{|\eta - \eta_{\text{ref}}|}{2} & \text{ if } |\eta - \eta_{\text{ref}}| \le 2 \\
   0 & \text{ otherwise } \\
  \end{cases}\;\text{.}
\end{eqnarray}
In this toy Monte Carlo simulation, no distinction on the $p_T$ or species of particles is used. 
When evaluating the per-trigger yield with correction methods, the symmetric $\eta$ acceptance, $[-2, 2]$, as in the previous PYTHIA simulation is used. Since new correction methods are not derived under this type of correlation, validity of the correction is not ensured. But this example will show the importance of finite-acceptance correction for several observables. 

\begin{figure}[tb!f]
\centering
\includegraphics[width=0.49\linewidth]{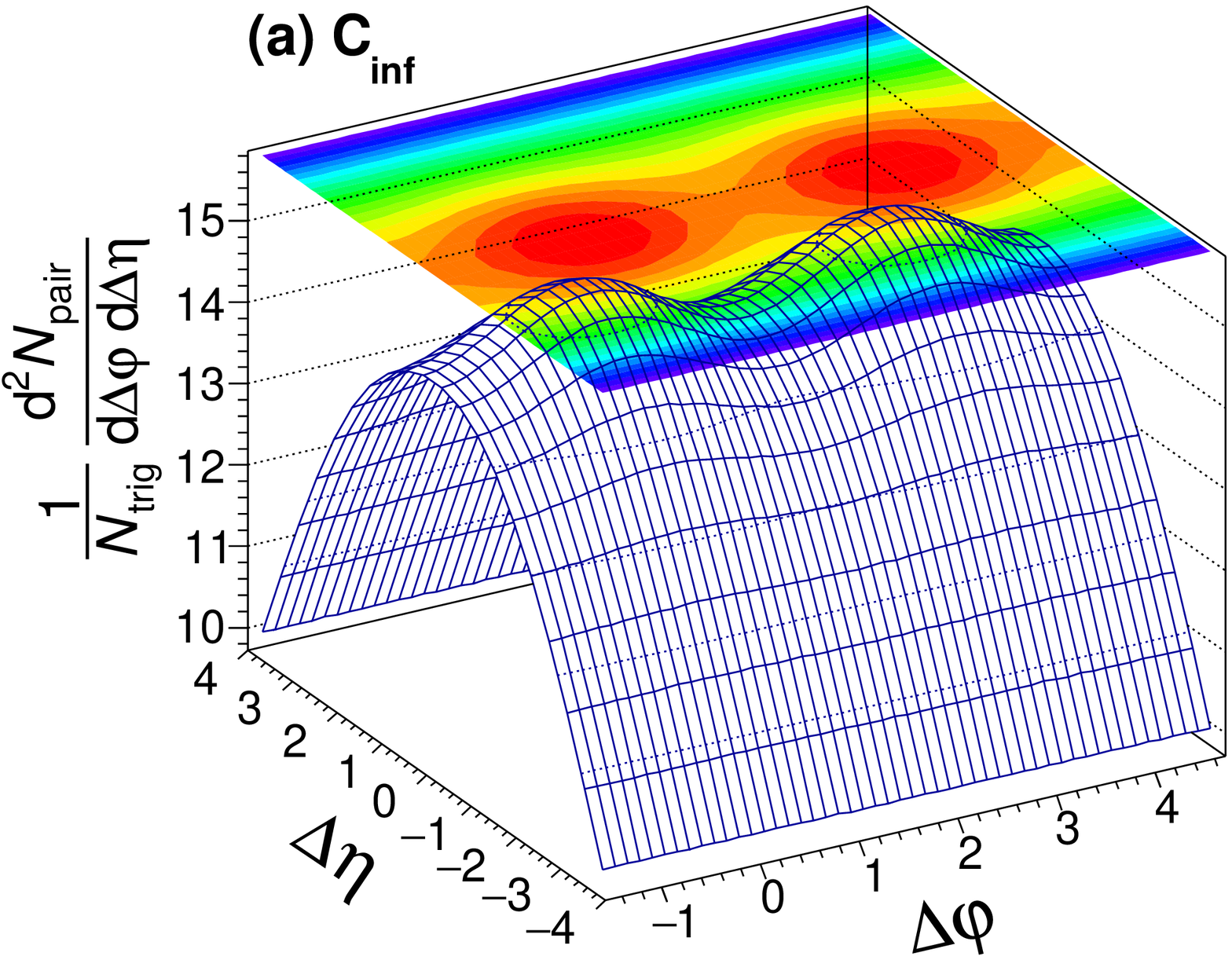}
\includegraphics[width=0.49\linewidth]{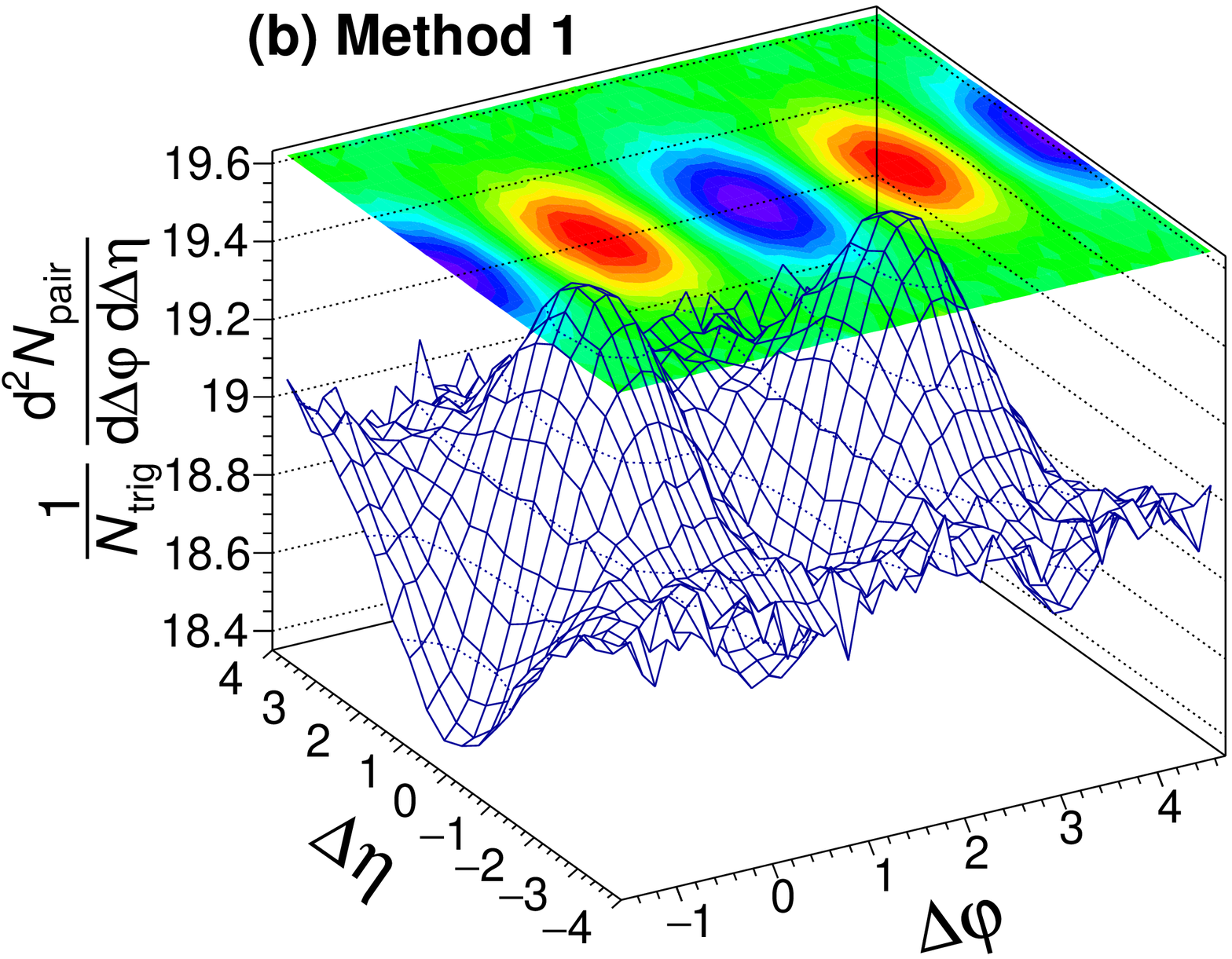}
\includegraphics[width=0.49\linewidth]{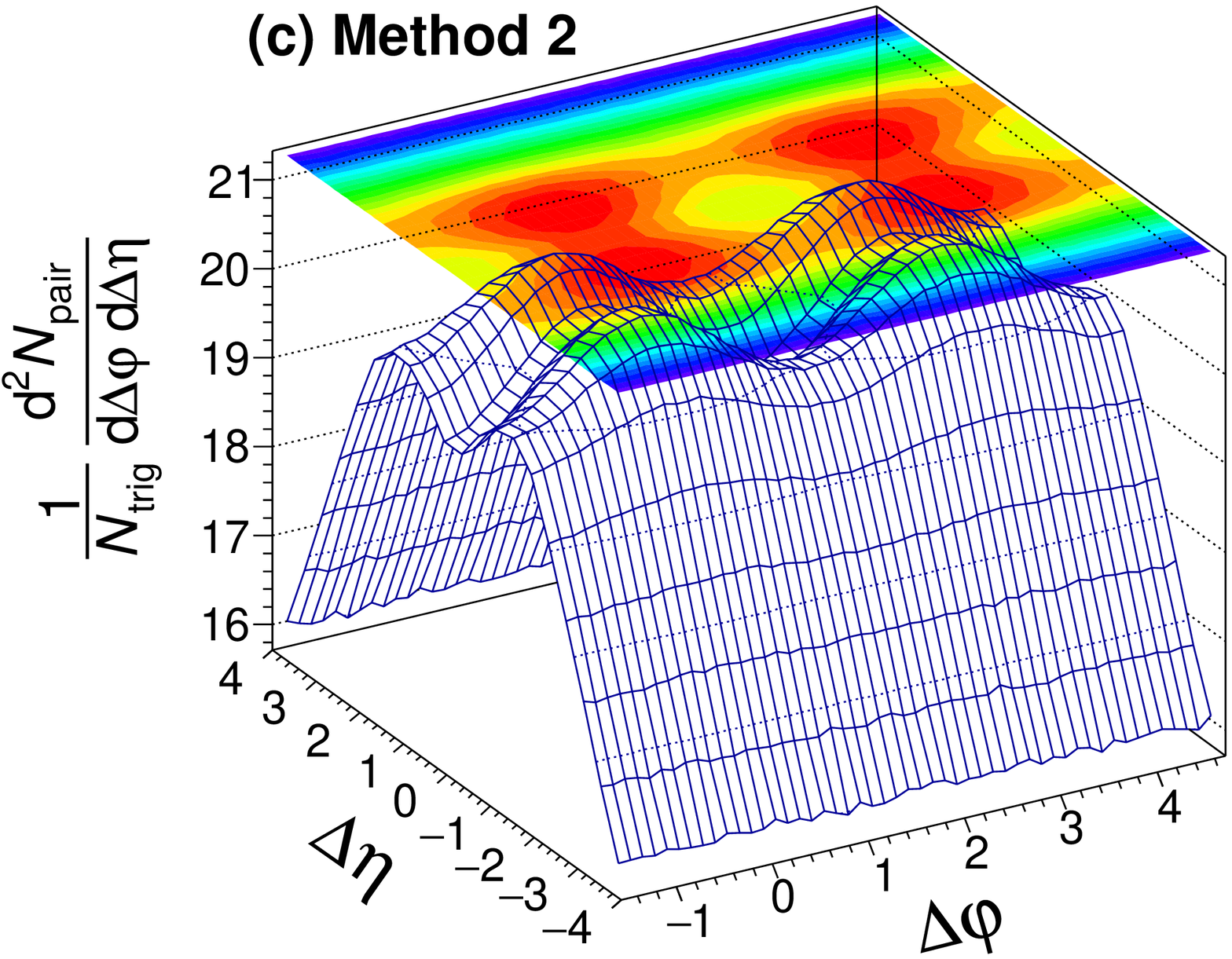}
\includegraphics[width=0.49\linewidth]{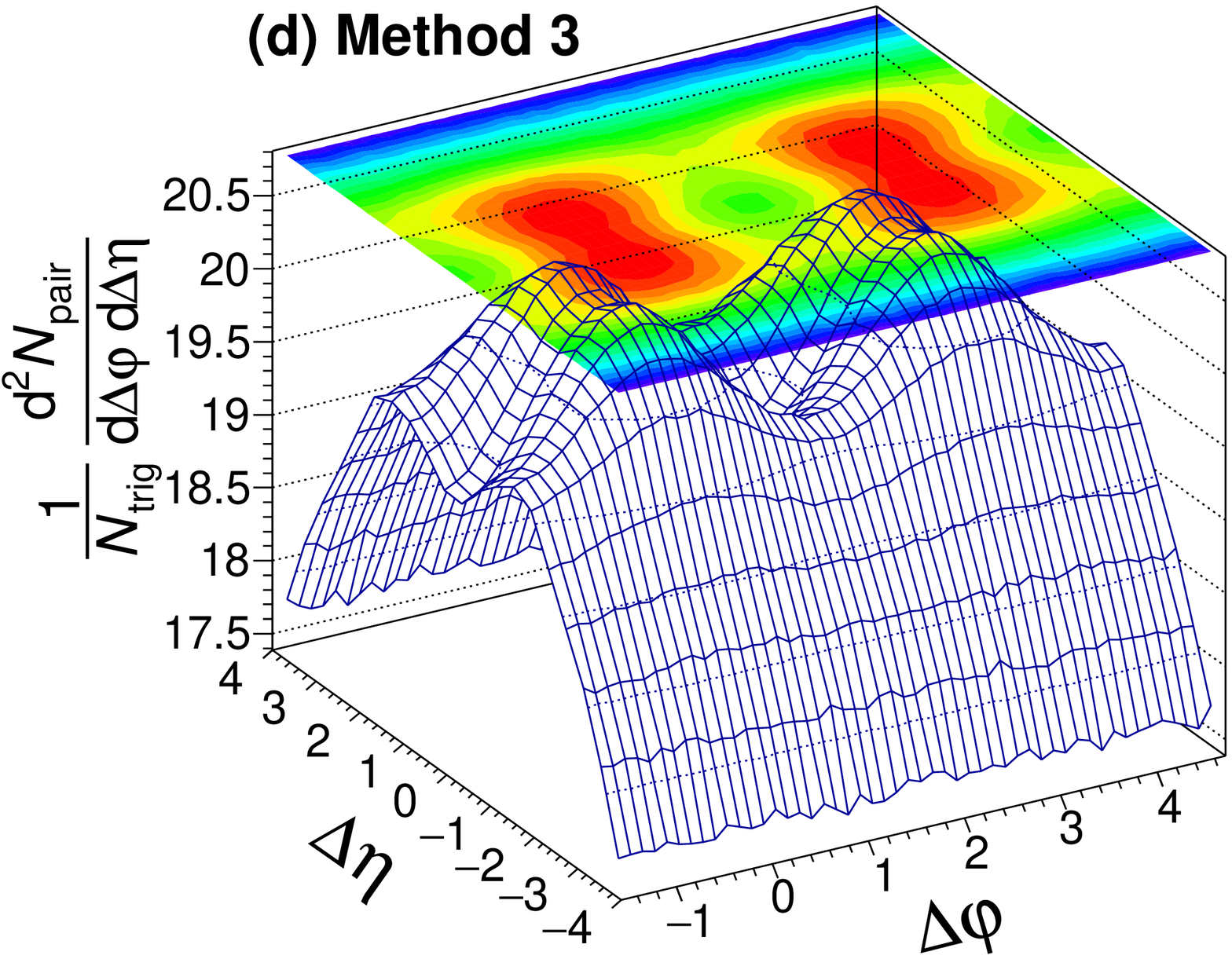}
\caption{\label{fig:flow_22_2D} (a)~Ideal per-trigger yield evaluated without $\eta$-cut; (b)~Per-trigger yield from Method~1; (c)~Method~2; (d)~Method~3; all with the $\eta$-acceptance $[-2,2]$.}
\end{figure}

Figure~\ref{fig:flow_22_2D} shows the per-trigger yields from infinite acceptance and three correction methods with finite acceptance. One obvious observation is that the per-trigger yield from Method~1 has larger relative yields at large $\Delta\eta$ than others.

The $v_2(\Delta\eta)$ extracted in each $\Delta\eta$ bin of the per-trigger yield is a relative quantity and does not depend on the finite-acceptance correction methods, since different methods only rescale the yield at each $\Delta\eta$ bin differently leaving the $\Delta\varphi$ shape intact.  It also means that $v_2(\Delta\eta)$ of the same-event function ($S(\Delta\varphi, \Delta\eta)$) without any finite-acceptance correction should be the same. However, if we consider integrated $v_2$, which is evaluated after projecting per-trigger yields to the $\Delta\varphi$ axis, a discrepancy occurs depending on the method. In our example, the per-trigger yield from Method~1 has larger yields at $|\Delta\eta|\sim4$, which corresponds to the $v_2 = 0$ region. Although $v_2$ fluctuates in event by event basis, the integrated $v_2$ will be smaller than one from the per-trigger yield with infinite acceptance. In other words, the yield of each $\Delta\eta$ bin works as a weighting factor for the projection, and if $v_2$ depends on $\Delta\eta$, the integrated $v_2$ should depend on the finite-acceptance correction method. Integrated $v_2$ from three correction methods still agree with the one from $C_{\text{inf}}$ by 11\% in our example. 

Some correlation analyses have not considered $(\Delta\varphi, \Delta\eta)$ space per-trigger yield, but have calculated $v_2$ based only on $\Delta\varphi$ space. This is equivalent to calculating $v_2$ from the same-event function, $S(\Delta\varphi, \Delta\eta)$, without any finite-acceptance correction and the integrated $v_2$ is dominated by the $v_2$ value at the $\Delta\eta$ bin with the largest yield of the same-event function. If $v_2$ varies depending on $\Delta\eta$, the integrated $v_2$ might be different from what is calculated from $\Delta\varphi$ space only.

\section{Summary and conclusion}
\label{Sec:Sum}

The commonly used method for calculating per-trigger normalized associated particle yield, which utilizes division by the normalized associated yields from mixed events, produces a normalized ratio function of correlated production and uncorrelated background~(Eq.~(\ref{CtrigR})). This ratio function differs from what is intended in the per-trigger yield, as it distorts the correlated signal shape and depends on the shape of the uncorrelated background. As a consequence, results of analyses using the $\Delta\varphi$ projection of per-trigger yield also depend on the shape of the uncorrelated background and may therefore differ from what they should be. To resolve this problem, we have discussed general correction procedure in $(x_{\text{t}}, \Delta x)$ space and derived new formulas for a correction in $\Delta x$ space that enable us to evaluate the per-trigger yield without dependence on the shape of the uncorrelated background. The formulas are derived under certain conditions and assumptions of correlated particle production, such as translational invariance of the signal, uniform correlated signal distribution~(Eq.~(\ref{Method2})) and jet-hadron-like correlations~(Eq.~(\ref{Method3})). They can be used as approximate corrections in more general cases. The validity of the new methods largely depends on the underlying mechanism that produces the correlated signals. We have tested the new methods using Monte Carlo simulations. A significant improvement with new methods was obtained in the case of an asymmetric signal distribution, (Fig.~\ref{fig:pythia_04_Ratio}) compared to the conventional method using event-mixing. The correction method with Eq.~(\ref{Method2}) is intended to be used in particles correlation analyses with midrapidity particles in symmetric nucleus-nucleus collisions where particles distribution is flat within the acceptance, while the other method with Eq.~(\ref{Method3}) is for jet-hadron or high $p_{\text{T}}$-triggered hadron-hadron two-particle correlations. 

It is difficult to precisely correct for finite-acceptance effects if the signal is a mixture of many physical mechanisms or if the type of the correlation is very different from the type that was assumed in the derivation of the new methods.

\section*{Acknowledgements}
The work of C.\ Loizides is supported in part by the U.S. Department of Energy, 
Office of Science, Office of Nuclear Physics, under contract number DE-AC02-05CH11231.
The work of S.\ Oh is supported by the US Department of Energy under Grant number DE-SC004168.
\bibliographystyle{utphys}
\bibliography{./biblio}{}

\begin{appendices}
\renewcommand{\theequation}{A-\arabic{equation}}	
\setcounter{equation}{0}  
\section{Derivations of correction formulas}  
\label{Sec:App}
In the case of a single common reference point between trigger and associated particle distributions, the derivation of Eq.~(\ref{Method2}) can be started from Eq.~(\ref{Nideal}), (\ref{Cideal}), (\ref{AccEq}), (\ref{1DCrEq}), and (\ref{1DNtrig}). If $g(X)$ is constant, it is possible to write as $g(X)=K$ with constant value $K$ over sufficiently large $x$ range. Then for $-\Delta_{\text{t}} < \Delta x < \Delta_{\text{t}}$, 
\begin{eqnarray}
N_{\text{trig}} &=& K\int_{-\infty}^{\infty}\int_{-\infty}^{\infty} f_{\text{t}}(x-X)\,A_{\text{t}}(x)\,\text{d}x\,\text{d}X \nonumber \\
&=& K\int_{-b}^{b}f_{\text{t}}(u)\,\text{d}u\,\int_{a_{\text{1,t}}}^{a_{\text{2,t}}}\text{d}t = (a_{\text{2,t}}-a_{\text{1,t}})\,K\int_{-b}^{b} f_{\text{t}}(u)\,\text{d}u\nonumber \\
&=& \Delta_{\text{t}}\,K\int_{-b}^{b} f_{\text{t}}(u)\,\text{d}u\; \text{,}
\end{eqnarray}
\begin{eqnarray}
C(\Delta x) &=& \frac{K}{N_\text{trig}}\int_{-\infty}^{\infty}\,\int_{-\infty}^\infty f_{\text{a}}(x-X-\Delta x)\,A_{\text{a}}(x-\Delta x)\,f_{\text{t}}(x-X)\,A_{\text{t}}(x)\,\text{d}x\,\text{d}X \nonumber \\
&=& \frac{K}{N_{\text{trig}}} \int_{-\infty}^{\infty}A_{\text{a}}(x-\Delta x)\,A_{\text{t}}(x)\,\text{d}x\int_{-\infty}^{\infty}f_{\text{a}}(u-\Delta x)\,f_{\text{t}}(u)\,\text{d}u \nonumber \\
&=& \frac{K}{N_{\text{trig}}}(A_{\text{a}}\star A_{\text{t}})\int_{-\infty}^{\infty}f_{\text{a}}(u-\Delta x)\,f_{\text{t}}(u)\,\text{d}u \nonumber \\
&=& \frac{(A_{\text{a}}\star A_{\text{t}})(\Delta x)}{\Delta_{\text{t}}}\frac{1}{\int_{-b}^{b} f_{\text{t}}(u)\,\text{d}u}\int_{-\infty}^{\infty} f_{\text{a}}(u-\Delta x)\,f_{\text{t}}(u)\,\text{d}u \nonumber \\
&=& \frac{(A_{\text{a}}\star A_{\text{t}})(\Delta x)}{\Delta_{\text{t}}}\,C_{\text{inf}}(\Delta x)\; \text{.}
\end{eqnarray}

Another condition that we can derive a formula with a single common reference point is when $f_{\text{t}}$ is a $\delta$-function, and we have Eq.~(\ref{Method3}). In this case, $f_{\text{t}}$ can be written as
\begin{equation}
f_{\text{t}}(x-X)=N\delta(x-X)\;{.}
\end{equation}
Then for the infinite acceptance case,
\begin{eqnarray}
(N_{\text{trig}})_{\text{inf}} &=& \int_{-\infty}^{\infty}g(X)\,\left(\int_{-\infty}^{\infty} N\delta_{\text{t}}(x-X)\,\text{d}x\right)\,\text{d}X \nonumber \\
& = & N\,\int_{-\infty}^{\infty}g(X)\,\text{d}X  \;\text{,}
\end{eqnarray}
\begin{eqnarray}
C_{\text{inf}}(\Delta x) &=& \frac{1}{(N_{\text{trig}})_{\text{inf}}}\int_{-\infty}^{\infty}\int_{-\infty}^{\infty}g(X)\,f_{\text{a}}(x-X-\Delta x)\,N\,\delta(x-X)\, \text{d}X\,\text{d}x \nonumber \\
&=&  \frac{N}{(N_{\text{trig}})_{\text{inf}}}\int_{-\infty}^{\infty}g(X)\, f_{\text{a}}(-\Delta x)\,\text{d}X \nonumber \\
&=& \frac{N}{N\,\int_{-\infty}^{\infty}g(X)\,\text{d}X}f_{\text{a}}(-\Delta x)\,\int_{-\infty}^{\infty}g(X)\,\text{d}X \nonumber \\
&=& f_\text{a}(-\Delta x)\;\text{.}
\end{eqnarray}
With finite acceptances as defined in Eq.~(\ref{AccEq}),
\begin{eqnarray}
N_{\text{trig}} &=& \int_{-\infty}^{\infty}\int_{-\infty}^{\infty} g(X)f_{\text{t}}(x-X)\,A_{\text{t}}(x)\,\text{d}X\,\text{d}x \nonumber \\
&=& N\int_{-\infty}^{\infty} g(x)\,A_{\text{t}}(x)\,\text{d}x \nonumber \\
&=& N\int_{a_{\text{1,t}}}^{a_{\text{2,t}}} g(x)\,\text{d}x\;\text{,}
\end{eqnarray}
\begin{eqnarray}
C(\Delta x) &=& \frac{N}{N_\text{trig}}\int_{-\infty}^{\infty}\text{d}X\int_{-\infty}^{\infty}\text{d}x\,g(X)\,f_{\text{a}}(x-X-\Delta x)\,A_{\text{a}}(x-\Delta x)\,\delta(x-X)\,A_{\text{t}}(x) \nonumber \\
&=& \frac{N}{N_\text{trig}}\int_{\infty}^{\infty}g(x)\,f_{\text{a}}(-\Delta x)\,A_{\text{a}}(x-\Delta x)\,A_{\text{t}}(x)\,\text{d}x \nonumber \\
&=& \frac{N}{N_\text{trig}}(A_{\text{a}}\star gA_{\text{t}})(\Delta x)\,f_{\text{a}}(-\Delta x) \nonumber \\
&=&\frac{(A_{\text{a}}\star N\,g\,A_{\text{t}})(\Delta x)}{N_{\text{trig}}}\,C_{\text{inf}}(\Delta x)\;{.}
\end{eqnarray}
Also, $N\,g(x)\,A_{\text{t}}(x)$ can be easily measured since
\begin{eqnarray}
n_{\text{trig}}(x)\,A_{\text{t}}(x) &=& \int_{-\infty}^{\infty}g(X)\,f_{\text{t}}(x-X)\,A_{\text{t}}(x)\,\text{d}X \nonumber \\
&=& N\int_{-\infty}^{\infty}g(X)\,\delta_{\text{t}}(x-X)\,A_{\text{t}}(x)\,\text{d}X \nonumber \\
&=& N\,g(x)\,A_{\text{t}}(x) \;\text{,}
\end{eqnarray}
where $n_{\text{trig}}(x)$ is a measured trigger particle distribution in $x$ within the acceptance. 
\\

For the case with two common reference correlation points, the derivation of Eq.~(\ref{Method2A}) can be started from Eq.~(\ref{2RefCtrig}), (\ref{2RefNtrig}), and (\ref{2RefCond}). Assuming  $g(X,Y) = G(X-Y)$ and $G(X-Y)$ has only values if $-c<X-Y<c$ with sufficiently large $c$ compared to $2b$, the range of $f_{\text{a}}$, the per-trigger yield with infinite acceptance becomes
\begin{eqnarray}
C_{\text{inf}}(\Delta x) &=& \frac{1}{N_\text{trig,inf}}\int_{-\infty}^{\infty}\text{d}x\int_{-\infty}^{\infty}dY\int_{-\infty}^{\infty}\text{d}X\,G(X-Y)\Big(f_{\text{a}}(x-Y-\Delta x)\,f_{\text{t}}(x-X)\nonumber \\
&&+f_{\text{t}}(x-Y)\,f_{\text{a}}(x-X-\Delta x)\Big) \nonumber \\
&=& \frac{2}{N_\text{trig,inf}}\int_{-\infty}^{\infty}\text{d}x\int_{-\infty}^{\infty}dY\int_{-\infty}^{\infty}\text{d}X\,G(X-Y)\,f_{\text{a}}(x-Y-\Delta x)\,f_{\text{t}}(x-X)\,\text{.}
\end{eqnarray}
We can divide the $\Delta x$ range into $\{\Delta x< -c-b\}$, $\{-c-b < \Delta x< -c+b\}$, $\{-c+b < \Delta x< c-b\}$, $\{c-b < \Delta x< c+b\}$ and $\{c+b < \Delta x\}$. Considering the assumption that $c$ is much larger than $b$, we are only interested in the $\{-c+b < \Delta x< c-b\}$ case. If $-c+b < \Delta x < c-b$,
\begin{eqnarray}
C_{\text{inf}}(\Delta x)&=& \frac{2}{N_\text{trig,inf}}\lim_{\beta \to +\infty}\int_{-\beta}^{\beta}\text{d}x\int_{-c}^{c}\text{d}u\int_{-b}^{b}\text{d}v\,G(u)\,f_{\text{t}}(v)\,f_{\text{a}}(u+v-\Delta x)\;\text{,}
\end{eqnarray}
\begin{eqnarray}
N_\text{trig,inf} &=& \int_{-\infty}^{\infty}\text{d}x\int_{-\infty}^{\infty}\text{d}X\int_{-\infty}^{\infty}dY\,G(X-Y)\Big( f_{\text{t}}(x-X) + f_{\text{t}}(x-Y)\Big) \nonumber \\
&=& 2\int_{-\infty}^{\infty}\text{d}x\int_{-\infty}^{\infty}\text{d}X\int_{-\infty}^{\infty}dY\,G(X-Y)f_{\text{t}}(x-X) \nonumber \\
&=& 2\lim_{\beta \to +\infty}\int_{-\beta}^{\beta}\text{d}x\int_{-c}^{c}\text{d}u\int_{-b}^{b}\text{d}v\,G(u)\,f_{\text{t}}(v)\;\text{.}
\end{eqnarray}
Considering $\beta\rightarrow\infty$ ,
\begin{eqnarray}
C_{\text{inf}}(\Delta x)&=& \frac{\int_{-c}^{c}\text{d}u\int_{-b}^{b}\text{d}v\,G(u)\,f_{\text{t}}(v)\,f_{\text{a}}(u+v-\Delta x)}{\int_{-c}^{c}\text{d}u\int_{-b}^{b}\text{d}v\,G(u)\,f_{\text{t}}(v)}\;\text{.}
\end{eqnarray}
With finite acceptances as defined in Eq.~(\ref{AccEq}) assuming $a \ll c$,
\begin{eqnarray}
N_\text{trig} &=&  \int_{-\infty}^{\infty}\text{d}x\int_{-\infty}^{\infty}\text{d}X\int_{-\infty}^{\infty}dY\,G(X-Y)\Big(f_{\text{t}}(x-X) + f_{\text{t}}(x-Y)\Big)\,A_{\text{t}}(x) \nonumber \\
&=& 2\int_{-\infty}^{\infty}\text{d}x\int_{-\infty}^{\infty}\text{d}X\int_{-\infty}^{\infty}dY\,G(x-Y)\,f_{\text{t}}(x-X)\,A_{\text{t}}(x) \nonumber \\
&=& 2(a_{\text{2,t}} - a_{\text{1,t}})\int_{-c}^{c}\text{d}u\int_{-b}^{b}\text{d}v\,G(u)\,f_{\text{t}}(v)\nonumber \\
&=& 2\,\Delta_{\text{t}}\int_{-c}^{c}\text{d}u\int_{-b}^{b}\text{d}v\,G(u)\,f_{\text{t}}(v)\;\text{,}
\end{eqnarray}
\begin{eqnarray}
C(\Delta x) &=& \frac{1}{N_\text{trig}}\int_{-\infty}^{\infty}\text{d}x\int_{-\infty}^{\infty}\text{d}Y\int_{-\infty}^{\infty}\text{d}X\,G(X-Y)\,A_{\text{a}}(x-\Delta x)\,A_{\text{t}}(x)\Big(\nonumber \\
&&f_{\text{a}}(x-Y-\Delta x)\,f_{\text{t}}(x-X)+f_{\text{t}}(x-Y)\,f_{\text{a}}(x-X-\Delta x)\Big) \nonumber \\
&=& \frac{2}{N_{\text{trig}}}\int_{-\infty}^{\infty}\text{d}x\int_{-\infty}^{\infty}\text{d}X\int_{-\infty}^{\infty}dY\,G(X-Y)\,f_{\text{a}}(x-Y-\Delta x)\,f_{\text{t}}(x-X)\,A_{\text{a}}(x-\Delta x)\,A_{\text{t}}(x) \nonumber \\
&=& \frac{2}{N_{\text{trig}}}\int_{-\infty}^{\infty}A_{\text{a}}(x-\Delta x)\,A_{\text{t}}(x)\,\text{d}x\int_{-c}^{c}\text{d}u\int_{-b}^{b}\text{d}v\,G(u)\,f_{\text{t}}(v)\,f_{\text{a}}(u+v-\Delta x) \nonumber \\
&=& \frac{2}{N_{\text{trig}}}A_{\text{a}}\star A_{\text{t}}\int_{-c}^{c}\text{d}u\int_{-b}^{b}\text{d}v\,G(u)\,f_{\text{t}}(v)\,f_{\text{a}}(u+v-\Delta x)  \nonumber \\
&=& \frac{(A_{\text{a}}\star A_{\text{t}})(\Delta x)}{\Delta_{\text{t}}}\frac{1}{\int_{-c}^{c}\text{d}u\int_{-b}^{b}\text{d}v\,G(u)\,f_{\text{t}}(v)}\int_{-c}^{c}\text{d}u\int_{-b}^{b}\text{d}v\,G(u)\,f_{\text{t}}(v)\,f_{\text{a}}(u+v-\Delta x) \nonumber \\
&=& \frac{(A_{\text{a}}\star A_{\text{t}})(\Delta x)}{\Delta_{\text{t}}}\,C_{\text{inf}}(\Delta x)\;\text{.}
\end{eqnarray}
\\

For the derivation of Eq.~(\ref{Method3A}), assuming $g(X,Y)= h(X)F(X-Y)$ and has only values if $-c<X-Y<c$ with sufficiently large $c$ compared to $2b$ and $f_{\text{t}}(x-X) = N\delta(x-X)$, 
\begin{eqnarray}
C_{\text{inf}}(\Delta x) &=& \frac{N}{N_\text{trig,inf}}\int_{-\infty}^{\infty}\text{d}x\int_{-\infty}^{\infty}dY\int_{-\infty}^{\infty}\text{d}X\,h(X)\,F(X-Y)\Big(f_{\text{a}}(x-Y-\Delta x)\,\delta(x-X)\nonumber \\
&&+\delta(x-Y)\,f_{\text{a}}(x-X-\Delta x)\Big) \nonumber \\
&=& \frac{2N}{N_\text{trig,inf}}\int_{-\infty}^{\infty}\text{d}X\int_{-\infty}^{\infty}\text{d}Y\,h(X)\,F(X-Y)\,f_{\text{a}}(X-Y-\Delta x)\,\text{.}
\end{eqnarray}
If $-c+b < \Delta x < c-b$,
\begin{eqnarray}
C_{\text{inf}}(\Delta x)&=& \frac{2N}{N_\text{trig,inf}}\lim_{\beta \to +\infty}\int_{-\beta}^{\beta}h(x)\,\text{d}x\int_{-b+\Delta x}^{b+\Delta x}F(u)\,f_{\text{a}}(u-\Delta x)\,\text{d}u\;\text{,}
\end{eqnarray}
\begin{eqnarray}
N_\text{trig,inf} &=& N\int_{-\infty}^{\infty}\text{d}x\int_{-\infty}^{\infty}\text{d}X\int_{-\infty}^{\infty}dY\,h(X)\,F(X-Y)\Big( \delta(x-X) + \delta(x-Y)\Big) \nonumber \\
&=& 2N\int_{-\infty}^{\infty}\text{d}X\int_{-\infty}^{\infty}\text{d}Y\,h(X)\,F(X-Y) \nonumber \\
&=& 2N\lim_{\beta \to +\infty}\int_{-\beta}^{\beta}h(x)\,\text{d}x\int_{-c}^{c}F(u)\,\text{d}u\;\text{.}
\end{eqnarray}
Considering $\beta\rightarrow\infty$ ,
\begin{eqnarray}
C_{\text{inf}}(\Delta x)&=& \frac{\int_{-b+\Delta x}^{b+\Delta x}F(u)\,f_{\text{a}}(u-\Delta x)\,\text{d}u}{\int_{-c}^{c}F(u)\,\text{d}u}\;\text{.}
\end{eqnarray}

With finite-acceptance effects, 
\begin{eqnarray}
N_\text{trig} &=&  N\int_{-\infty}^{\infty}\text{d}x\int_{-\infty}^{\infty}\text{d}X\int_{-\infty}^{\infty}dY\,h(X)\,F(X-Y)\Big(\delta(x-X) + \delta(x-Y)\Big)\,A_{\text{t}}(x) \nonumber \\
&=& 2N\int_{-\infty}^{\infty}\text{d}X\int_{-\infty}^{\infty}\text{d}Y\,h(X)\,F(X-Y)\,A_{\text{t}}(X) \nonumber \\
&=& 2N\int_{a_{\text{1,t}}}^{a_{\text{2,t}}}h(x)\,\text{d}x\int_{-c}^{c}\text{d}u\,F(u)\,\text{,}
\end{eqnarray}
\begin{eqnarray}
C(\Delta x) &=& \frac{N}{N_\text{trig}}\int_{-\infty}^{\infty}\text{d}x\int_{-\infty}^{\infty}\text{d}Y\int_{-\infty}^{\infty}\text{d}X\,h(X)\,F(X-Y)\,A_{\text{a}}(x-\Delta x)\,A_{\text{t}}(x)\Big(\nonumber \\
&&f_{\text{a}}(x-Y-\Delta x)\,\delta(x-X)+\delta(x-Y)\,f_{\text{a}}(x-X-\Delta x)\Big) \nonumber \\
&=& \frac{2N}{N_{\text{trig}}}\int_{-\infty}^{\infty}\text{d}X\int_{-\infty}^{\infty}\text{d}Y\,h(X)\,F(X-Y)\,f_{\text{a}}(X-Y-\Delta x)\,A_{\text{a}}(X-\Delta x)\,A_{\text{t}}(X) \nonumber \\
&=& \frac{2N}{N_{\text{trig}}}\int_{-\infty}^{\infty}h(X)\,A_{\text{a}}(X-\Delta x)\,A_{\text{t}}(X)\,\text{d}X\int_{-b+\Delta x}^{b+\Delta x}F(u)\,f_{\text{a}}(u-\Delta x)\,\text{d}u \nonumber \\
&=& \frac{2N}{N_{\text{trig}}}(A_{\text{a}}\star h\,A_{\text{t}})\int_{-b+\Delta x}^{b+\Delta x}F(u)\,f_{\text{a}}(u-\Delta x)\,\text{d}u \nonumber \\
&=& \frac{(A_{\text{a}}\star h\,A_{\text{t}})(\Delta x)}{\int_{a_{\text{1,t}}}^{a_{\text{2,t}}}h(x)\,\text{d}x}\frac{1}{\int_{-c}^{c}F(u)\,\text{d}u}\int_{-b+\Delta x}^{b+\Delta x}F(u)\,f_{\text{a}}(u-\Delta x)\,\text{d}u \nonumber \\
&=& \frac{(A_{\text{a}}\star h\,A_{\text{t}})(\Delta x)}{\int_{a_{\text{1,t}}}^{a_{\text{2,t}}}h(x)\,\text{d}x}\,C_{\text{inf}}(\Delta x)\nonumber \\
&=& \frac{(A_{\text{a}}\star n_{\text{trig}}A_{\text{t}})(\Delta x)}{N_{\text{trig}}}\,C_{\text{inf}}(\Delta x)\,\text{.}
\end{eqnarray}
\end{appendices}

\end{document}